\documentclass{emulateapj}
\usepackage{graphicx}
\usepackage{times}

\begin{document}

\markboth{Cheng-Jiun Ma} {Spectroscopic diagnostics in MACSJ0717}
\title{The spatial distribution of galaxies of different spectral types in the massive intermediate-redshift
  cluster MACSJ0717.5+3745\footnote{Based in part on data collected at Subaru
    Telescope, which is operated by the National Astronomical Observatory of
    Japan. Based also, in part, on observations obtained with MegaPrime/MegaCam,
    a joint project of CFHT and CEA/DAPNIA, at the Canada-France-Hawaii
    Telescope (CFHT) which is operated by the National Research Council (NRC) of
    Canada, the Institute National des Sciences de l'Univers of the Centre
    National de la Recherche Scientifique of France, and the University of
    Hawaii. The spectroscopic data presented herein were obtained at the
    W.M.\ Keck Observatory, which is operated as a scientific partnership among
    the California Institute of Technology, the University of California and the
    National Aeronautics and Space Administration. The Observatory was made
    possible by the generous financial support of the W.M. Keck Foundation.}}
\author{Cheng-Jiun Ma, Harald Ebeling, David Donovan, and Elizabeth Barrett}
\affil{Institute for Astronomy, University of Hawaii, 2680 Woodlawn Drive, Honolulu, HI 96822, USA}

\begin{abstract}

We present the results of a wide-field spectroscopic analysis of the galaxy
population of the massive cluster MACSJ0717.5+3745 and the surrounding
filamentary structure (z$=$0.55), as part of our systematic study of the 12 most
distant clusters in the MACS sample. Of 1368 galaxies spectroscopically observed
in this field, 563 are identified as cluster members; of those, 203 are
classified as emission-line galaxies, 260 as absorption-line galaxies, and 17 as
E+A galaxies (defined by $\frac{H_{\delta}+H_{\gamma}}{2}>6$\AA\ and no
detection of [OII] and $H_{\beta}$ in emission). The variation of the fraction
of emission- and absorption-line galaxies as a function of local projected
galaxy density confirms the well-known morphology-density relation, and becomes
flat at projected galaxy densities less than $\sim 20$Mpc$^{-2}$. Interestingly,
16 out of 17 E+A galaxies lie (in projection) within the ram-pressure stripping
radius around the cluster core, which we take to be direct evidence of
ram-pressure stripping being the primary mechanism that terminates
star-formation in the E+A population of galaxy clusters. This conclusion is
supported by the rarity of E+A galaxies in the filament which rules out galaxy
mergers as the dominant driver of evolution for E+A galaxies in clusters. In
addition, we find the 42 e(a) and 27 e(b) member galaxies, i.e., the
dusty-starburst and starburst galaxies respectively, to be spread out across
almost the entire study area. Their spatial distribution, which shows a strong
preference for the filament region, suggests that starbursts are triggered in
relatively low-density environments as galaxies are accreted from the field
population.

\end{abstract}


\section{Introduction}

It is widely accepted that the effect of environment is one of the most
important factors for galaxy evolution. Unlike field galaxies for which the
dominant physical mechanism driving evolution is galaxy-galaxy mergers, as has
been revealed in many large-scale surveys \citep{gome03,lewi03,coop07},
cluster galaxies evolve in a more complex fashion (see, e.g., the review by
\citet{bowe04}). As a result, cluster galaxies behave differently from field
galaxies at the same epoch. One of the best-studied examples of this difference
between the properties of galaxies in clusters and in the field is the
morphology-density relation\citep{oeml74,dres80}, i.e., the fact that the
fraction of late-type galaxies drops precipitously toward the center of a
cluster, whereas the fraction of early-type galaxies increases dramatically. In
addition, the Butcher-Oemler effect\citep{bo78}, which states that galaxies in
rich clusters tend to be bluer and more active (i.e.\ feature emission lines)
with increasing cluster redshift out to at least $z\sim$ 0.8, implies that the
build-up of the morphology-density relation is still ongoing at redshifts less
than unity. Therefore, much effort \citep{couc87,dres97, pogg99, smit05,
  post05,mora06,tran07} has gone into establishing whether and, if so, how blue
late-type galaxies are transformed into early-type galaxies, specifically S0
galaxies.
 
Clusters at intermediate redshift ($z\sim 0.5$) provide us with a unique
opportunity to examine this hypothesized transformation as well as the dominant
physical mechanisms driving it. Recent work in this field includes a study of
the morphological distribution of galaxies in the cluster Cl0024+16 at $z=0.39$
\citep{treu03}, and a comparison of the spatial distribution and spectroscopic
properties of spiral galaxies with quenched star formation (passive spirals) and
newly formed S0 galaxies in, again Cl0024+16, as well as in the more massive
system MS0451.6--0305 at $z=0.55$ \citep{mora07}. The authors of these studies
suggest that ram-pressure stripping \citep{gunn72} is not the only mechanism
involved in building up the population of S0 galaxies from late-type galaxies,
although it does accelerate the process. The transformation is actually
initiated much farther away from the cluster center, and at a slower pace, by
the processes of strangulation \citep{lars80} or galaxy-galaxy harassment
\citep{moor96}. Additional evidence of this evolution from blue late-type
galaxies to S0 galaxies is presented by \citet{tran07} who, using the
spectroscopic diagnostics H$_{\delta}$ and $D_n(4000)$ \citep{kauf03} in their
study of the cluster MS1054.4--0321 ($z=0.823$), confirms that S0 galaxies in
the cluster are, in general, younger than the elliptical
galaxies. \citet{tran07} suggest that these S0 galaxies have recently been
converted from late-type galaxies falling in from the field (see also
\citet{tran05}). 

To improve our understanding of the different physical effects governing galaxy
evolution from the cluster outskirts to the cluster core, we have embarked on an
extensive magnitude-limited wide-field spectroscopic survey of the galaxy
population of 12 clusters with redshifts above $0.5$, selected from the MAssive
Cluster Survey (MACS) \citep{ebel04,ebel07}. In this paper we present results
from our pilot study of MACSJ0717.5+3745 ($z=0.545$). MACSJ0717.5+3745, which
has the most extensive spectroscopic data set of these 12 at present, features a
complex X-ray morphology in the cluster core as well as a giant filament
described by \citet{ebel04}. We here focus on the spectroscopic analysis and
classification of galaxies in MACSJ0717.5+3745, with special emphasis on E+A
galaxies, a galaxy type that, as shall be explained in the following section,
holds great promise as a diagnostic for cluster-related mechanisms of galaxy
evolution.

This paper is organized as follows. In \S2, we review the properties of E+A
galaxies. In \S3, we describe the data used for this study and give an overview
of the data reduction procedure. In \S4 we discuss the selection of the cluster
members used in our analysis, and describe the equivalent-width measurement and
spectral classification. In \S5, we present results based on the spatial
distribution of each spectral type of galaxies. In \S6 we interpret our
findings; a conclusion is given in \S7.  Throughout this paper, we adopt the
concordance $\Lambda$CDM cosmology with $h_0=0.7$, $\Omega_{\lambda}=0.7$,
$\Omega_m = 0.3$. Magnitudes are quoted in the AB system. 

\section{E+A Galaxies}
One of the most promising approaches to understanding how blue and active
late-type galaxies are transformed into red and passive early-type systems
\citep{bowe04} is through studies of galaxies caught in transition between the
two types, for example E+A galaxies \citep{dres83}, also called k+a, a+k, or
post-starburst galaxies. Although there is no unique definition of this
transition type, its representatives are characterized by the presence of strong
Balmer absorption lines in their spectra, indicating a large fraction of A-type
stars, and weak or no ongoing star formation, often quantified by the strength
of the [OII] emission line. Since the duration of the E+A phase is relatively
short (less than 1 Gyr) E+A galaxies should be good tracers of the environment
that is host to a critical part of the evolution of galaxies in clusters.

Although E+A galaxies were first found in clusters at intermediate redshift
(e.g., Dressler \& Gunn 1983, Couch \& Sharples 1987), their origin and habitat
was soon proven to be more varied by the discovery that the majority of E+A
galaxies in the local universe are in fact not located in dense environments
\citep{zabl96,goto03,goto07, hogg06, quin04}. As for the possible physical
origin of this type of galaxies, \citet{zabl96} and \citet{yang04} report
that many of the E+A galaxies in the field show evidence of galaxy mergers. At
higher redshift, however, E+A galaxies are predominantly found in the much
denser environment of clusters \citep{tran03,tran04}, which have thus been
suspected of being instrumental in the termination of star-formation activity in
these systems. In addition, \citet{pogg04} show that, at M$_v > -18.5$, the
post-starburst galaxies in Coma are fainter than the typical E+A galaxies found
in clusters at higher redshift, which suggests that the evolution of
post-starburst galaxies follows the "downsizing" trend proposed by
\citet{cowie96}. Any discrepancies between the properties of E+A galaxies in the
local field and in intermediate-redshift clusters may thus be explained by the
combination of evolution and selection biases, in the sense that the local
surveys sample predominantly luminous galaxies and hence do not probe the
population of fainter E+A galaxies in clusters (the faintest E+A galaxy in
\citet{zabl96} features M$_R \sim -18.5$, whereas the magnitude limit in
\citet{goto03} is M$_r < -21.84$).

Unlike previous studies \citep{tran03, tran04, goto03}, the goal of this work is
not the issue of whether E+A galaxies are more likely to be found in clusters or
in the field; rather, we focus on the spatial distribution of E+A galaxies
around the massive cluster MACSJ0717.5+3745 to isolate the most probable
physical mechanism responsible for their creation in a cluster environment.  In
this context we also attempt to address the question of the origin of the E+A
population by investigating the spatial distribution of possible star-burst
galaxies, the so-called e(a) and e(b) galaxies \citep{dres99,pogg00}.

\section{Data and Analysis}

Using optical images obtained with the Suprime-Cam wide-field camera on the
Subaru 8m telescope on Mauna Kea \citep{miya02}, we identify galaxies within the
$34\times 27$ arcmin$^2$ Suprime-Cam field of view. We then employ color
selection to obtain spectra of a subset of presumed cluster members, primarily
using the DEep Imaging Multi-Object Spectrograph (DEIMOS) on the Keck-II 10m
telescope. In order to study the spectroscopic properties of galaxies in
different environments, the DEIMOS masks are designed to cover the entire
cluster including the filament and a satellite cluster. We use the spectroscopic
data to determine the redshift of galaxies, as well as their spectral types. We
follow the spectral classification scheme of \citet{tran03} (with some
modifications) to categorize the cluster galaxies into emission-line,
absorption-line, and E+A galaxies.  The emission-line galaxies can be further
subdivided into e(a), e(b), and e(c) galaxies according to the equivalent width
of the H$_{\delta}$ absorption and [OII] emission lines
\citep{dres99}. Complementing our optical analysis of the galaxy population of
this system, X-ray data obtained with the imaging array of the Chandra Advanced
CCD Imaging Spectrometer (ACIS-I) are analyzed to provide a basic model of the
gaseous intracluster medium.

\subsection{Optical Photometry}
We use images in the B, V, R$_c$, I$_c$, and z$^\prime$ bands obtained with
Suprime-Cam, supplemented by images in the u$^*$ band obtained with the
MegaPrime camera on the CFHT 3.6m telescope. The observations were performed
from December 2000 to February 2004. All data are reduced employing standard
techniques, which have, however, been adapted to deal with the special
characteristics of the Suprime-Cam and MegaPrime data; details are given by
\citet{dono07}.

In order to allow a robust estimate of the spectral energy distribution (SED) to
be obtained for all objects within the field of view, the imaging data from the
various passbands are seeing-matched using the technique described in
\citet{jeyh08}; this ensures that the images for all bands have the same
effective spatial resolution of 1$\arcsec$.

Object catalogues are then created with SExtractor version 2.4.3\citep{bertin96}
in ``dual image'' mode with the R-band image as the reference detection
image. We separate stars from galaxies by fitting the stellar sequence in the
distributions of magnitude vs peak surface-brightness, and magnitude vs
half-light radius. Isophotal galaxy magnitudes calculated for elliptical
apertures are returned by SExtractor in the MAG\_AUTO parameter. The photometric
calibration of the Suprime-Cam imaging data is performed by means of a snapshot
observation of a nearby SDSS field\footnote{The SDSS photometry is transformed
  into the Johnson-Cousin system using the equation of Lupton (2005) provided on
  http://www.sdss.org/dr5/algorithms/sdssUBVRITransform.html.} and uses hundreds
of stars with magnitudes ranging from 16 mag to 18 mag. The calibration of our
MegaCam data is supplied by CFHT's Elixir
project\footnote{http://www.cfht.hawaii.edu/Instruments/Elixir/}.

\subsection{X-ray Data}  
\begin{figure}[h]
\epsscale{1.0} 
\plotone{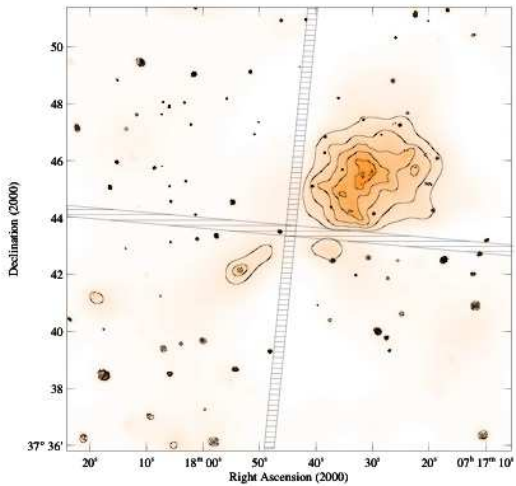} 
\figcaption[paper_xray.ps]{X-ray
  emission in the field of MACSJ0717.5+3745 as observed with Chandra's ACIS-I
  detector in the 0.5--7 keV band. The shown image has been adaptively smoothed
  using the \textit{asmooth} algorithm \citep{ebel06} requiring a minimal
  significance with respect to the local background of 99\%. The shaded cross
  marks the regions falling onto the chip gaps of the ACIS-I detector. A
  close-up view of the cluster core based on the same X-ray data has been
  previously published by \citet{ebel07}. [The full resolution figure will be published in ApJ] \label{xrayim}}
\end{figure}

The X-ray data are used to derive global cluster properties, most importantly
the virial radius and the ram-pressure stripping radius.  To this end, we
analyze the data taken with the Chandra ACIS-I instrument in January 2003 with a
total exposure time of 60 ks (ObsID 4200). The data reduction is performed
following the procedure described by \citet{rude05}. Fig.~\ref{xrayim} shows the
adaptively smoothed X-ray emission in the 0.5--7 keV band as observed by ACIS-I.
The X-ray luminosity and temperature are listed in Table~\ref{xrayprop}; a
dynamical analysis based on the X-ray properties and the velocity dispersion
will be presented in \S5.3.

\subsection{Optical Spectroscopy}

MACSJ0717.5+3745 has the most extensive spectroscopic data in our on-going
follow-up study of the dynamical structure and galaxy properties in 12 clusters
with $z>0.5$ in the MACS sample \citep{ebel07,jeyh08}. The spectroscopic data
set used here is compiled primarily from the observation of 18 masks with the
DEIMOS spectrograph on the Keck-II telescope. In addition, some spectra obtained
using the Low Resolution Image Spectrometer (LRIS) on the Keck-I telescope, and
the Gemini Multi-Object Spectrographs (GMOS) on the Gemini telescope are
included. The instrumental configuration of these observations is summarized by
\citet{barrett05}, from where we have extracted the essential information listed
in Table~\ref{spectdata}. The chosen setup is a compromise between the
requirements of wavelength coverage from the [OII] $\lambda\lambda 3727$ line to
the H$_{\beta}$ line at a redshift of about 0.5, and moderately high spectral
resolution. After reducing the spectra using the standard DEIMOS pipeline
developed by the DEEP2 team, the redshifts are determined and verified manually
using at least two prominent spectral features, such as (in absorption) Calcium
H and K, H$_{\delta}$, or the G band, and (in emission) [OII] $\lambda\lambda
3727$, H$_{\beta}$, and [OIII] $\lambda\lambda 4959,5007$. Based on repeated
observations of individual objects, we estimate the error in the resulting
redshifts to be about 70 km/s, which is consistent with the spectral resolution
of the observations.

\begin{figure*}[ht!]
\plotone{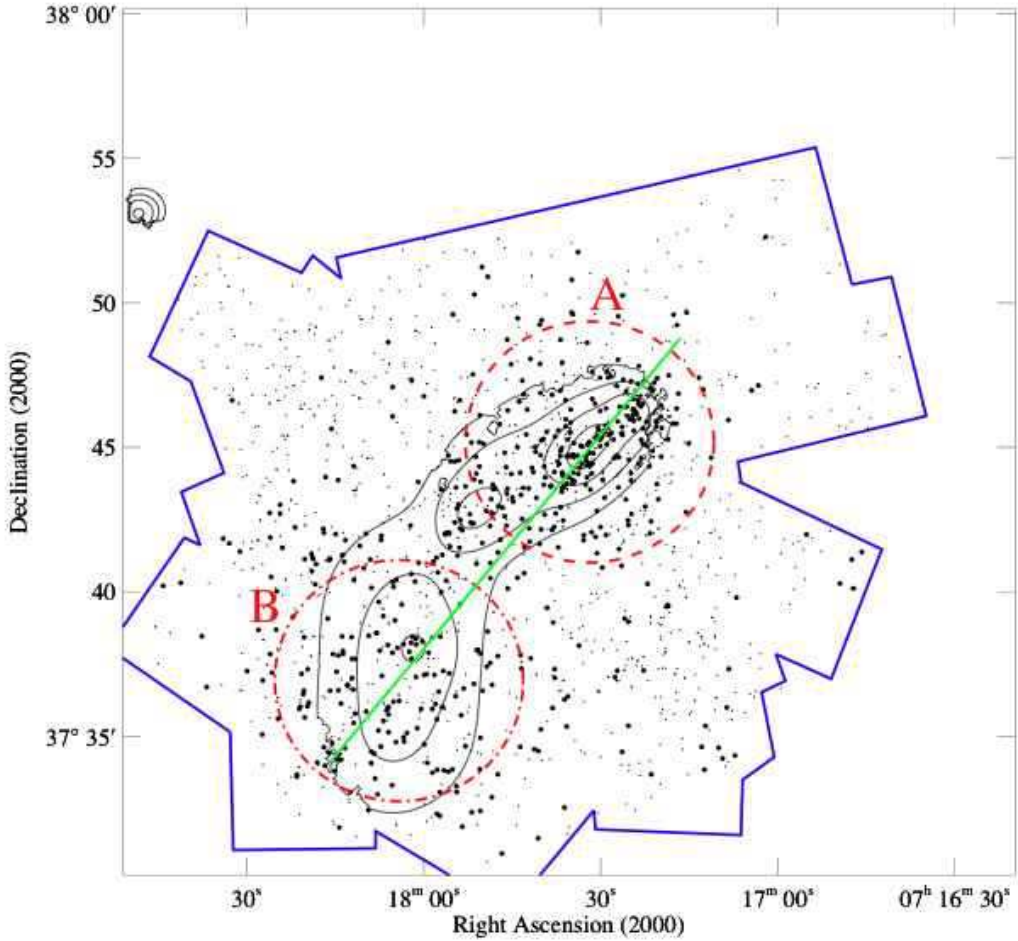}
 \figcaption[f2.eps]{Spatial distribution of galaxies
  observed spectroscopically: small dots mark the location of all sources with
  measured spectrum, large dots mark confirmed cluster members. The overlaid
  contours show the projected galaxy density as computed by
  \textit{asmooth}\citep{ebel06}. An image of the distribution of all galaxies
  in the photometric redshift catalogue (\S5.1) is adaptively smoothed such that
  all features in the smoothed image are $3\sigma$ significant with respect to
  the local background, where feasible. The plotted contours are located at
  $17$, $35$, $71$, $142$, and $285$ Mpc$^{-2}$, respectively; they are truncated
  where the significance of the signal is less than $3\sigma$. The blue boundary
  indicates the outline of all DEIMOS masks and defines the study region for
  this paper. All areas observed with GMOS and LRIS are covered by the DEIMOS
  masks. The two red circles (1.6 Mpc radius) indicate two regions of particular
  interest which will be discussed in more detail in \S5.4 and \S5.6: (a) the
  center of the cluster (dashed), and (b) the endpoint of the filament
  (dash-dotted). Both of these regions contain a local peak of the galaxy
  density. The green line crudely describes the large-scale orientation of the
  overall system and will be used in \S5.3 to assess how galaxy properties vary
  as a function of environment.[The full resolution figure will be published in ApJ]\label{plain}}
\end{figure*}

\begin{figure}[ht]
\epsscale{1.0} \plotone{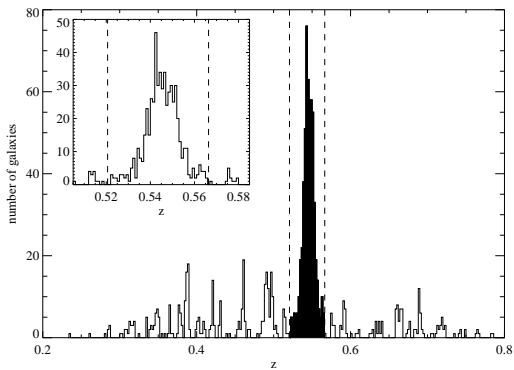} \figcaption[f3.eps]{Redshift
  distribution for all spectroscopically observed galaxies. The systemic
  redshift of MACSJ0717.5+3745 is found to be $z_{\rm cl}=0.545$. The dashed
  lines at $z=0.522$ and $z=0.566$ ($z=z_{\rm cl}\pm3\sigma$) mark the
  spectroscopic extent of the cluster; the black area highlights the cluster
  members.  A zoomed-in view of the redshift range near the cluster redshift is
  shown in the sub-panel.\label{redshiftplt}}
\end{figure}

The final spectroscopic catalogue comprises 1103 galaxies, of which 563 are
cluster members (Table~\ref{spectsum}). Fig.~\ref{plain} shows the location of
the spectroscopically observed objects within our study region which is
highlighted by the outline of the DEIMOS masks. The full redshift distribution
is shown in Fig.~\ref{redshiftplt}. Cluster membership is assigned to all
galaxies with $0.522<z<0.566$, the redshift range given by $z=z_{cl}\pm
2.5\sigma$, where the cluster redshift z$_{cl}=0.5446$ and velocity dispersion
$\sigma=1612$ km s$^{-1}$ (Table~\ref{xrayprop}) are determined from the
redshift distribution of galaxies within the virial radius of the
cluster\citep{ebel07}. The final catalogue, including redshifts and equivalent
widths of the spectral features ([OII], H$_{\delta}$, H$_{\gamma}$, D$_n$(4000),
H$_{\beta}$, and [OIII]), will be published in a separate data paper.

\begin{deluxetable}{lccccc}
\tabletypesize{\scriptsize}
\tablewidth{0pc}
\tablecolumns{6} 
\tablecaption{Summary of spectroscopic observations \label{spectdata}}
\tablehead{ 
\colhead{Instrument} & \colhead{Date} & \colhead{Grating} & \colhead{$\lambda_c$} & \colhead{Filter} &\colhead{Total Exposure Time} \\
\colhead{}                    &  \colhead{}         & \colhead{}              & \colhead{(\AA)}              & \colhead{}          &\colhead{(sec)} 
}
\startdata
LRIS\tablenotemark{a}       & 2000-11-20 & 600/7500 & 6200 & GC495  & 3600 \\
LRIS\tablenotemark{a}       & 2000-11-20 & 600/7500 & 6200 & GC495  & 4800 \\
LRIS\tablenotemark{a}       & 2000-11-21 & 600/7500 & 6200 & GC495  & 3600 \\
LRIS\tablenotemark{a}       & 2002-11-29 & 600/7500 & 6500 & GC495  & 5400 \\
LRIS\tablenotemark{a}       & 2002-11-29 & 600/7500 & 6500 & GC495  & 6000 \\
GMOS-N\tablenotemark{a}     & 2004-03-12 & B600     & 6700 & GC455  & 5400 \\
DEIMOS\tablenotemark{a}     & 2003-12-23 & 600ZD    & 6500 & GC455  & 7200 \\
DEIMOS\tablenotemark{a}     & 2004-12-16 & 600ZD    & 7000 & GC455  & 5400\\
DEIMOS\tablenotemark{a}     & 2004-12-16 & 600ZD    & 7000 & GC455  & 5400\\
DEIMOS\tablenotemark{a}     & 2004-12-16 & 600ZD    & 7000 & GC455  & 5400\\
DEIMOS\tablenotemark{a}     & 2005-02-12 & 600ZD    & 7000 & GC455  & 3239 \\
DEIMOS\tablenotemark{a}     & 2005-02-12 & 600ZD    & 7000 & GC455  & 6900 \\
\cline{1-6} \vspace*{-2mm} \\
DEIMOS     & 2006-12-22 & 600ZD    & 6300 & GC455  & 5400 \\
DEIMOS     & 2006-01-31 & 600ZD    & 6300 & GC455  & 6047 \\
DEIMOS     & 2006-01-31 & 600ZD    & 6300 & GC455  & 5400 \\
DEIMOS     & 2006-01-31 & 600ZD    & 6300 & GC455  & 5400 \\
DEIMOS     & 2006-01-31 & 600ZD    & 6300 & GC455  & 4991 \\
DEIMOS     & 2006-02-01 & 600ZD    & 6300 & GC455  & 5400 \\
DEIMOS     & 2006-02-01 & 600ZD    & 6300 & GC455  & 3600 \\
DEIMOS     & 2006-02-01 & 600ZD    & 6300 & GC455  & 5400 \\
DEIMOS     & 2006-02-01 & 600ZD    & 6300 & GC455  & 3600 \\
DEIMOS     & 2008-01-05 & 600ZD    & 6300 & GC455  & 5400 \\
DEIMOS     & 2008-01-06 & 600ZD    & 6300 & GC455  & 5400 \\
DEIMOS     & 2008-01-07 & 600ZD    & 6300 & GC455  & 5400
\enddata

\tablenotetext{a} {Reference: \citet{barrett05}}

\end{deluxetable}

\section{Spectroscopic Analysis}
\subsection{Object Selection and Completeness}

\begin{figure}[ht]
  \epsscale{1.0} \plotone{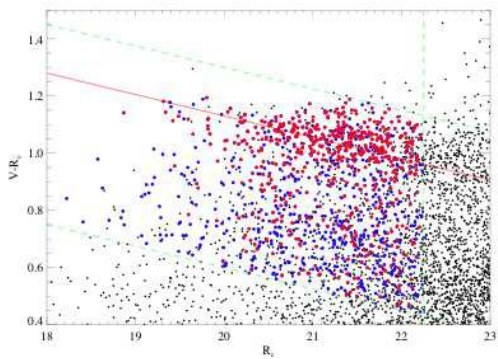} \figcaption[f4.eps]{Color magnitude diagram
    for galaxies inside our study region as defined in Fig~\ref{plain}. Blue
    filled circles mark objects with observed spectra; red filled circles denote
    spectroscopically confirmed cluster members (cf.\
    Fig.~\ref{redshiftplt}). Black dots are all remaining objects in our
    SExtractor galaxy catalogue. The dashed lines indicate the selection
    criteria used for our spectroscopic survey. The red sequence marked by the
    solid line is obtained from a linear fit to the data points redder than
    V-R$_{c}\sim 0.9$ and brighter than R$_{c}\sim 22.5$. [The full resolution figure will be published in ApJ]\label{cmd}}
\end{figure}

\begin{figure}[h]
\epsscale{1.0} \plotone{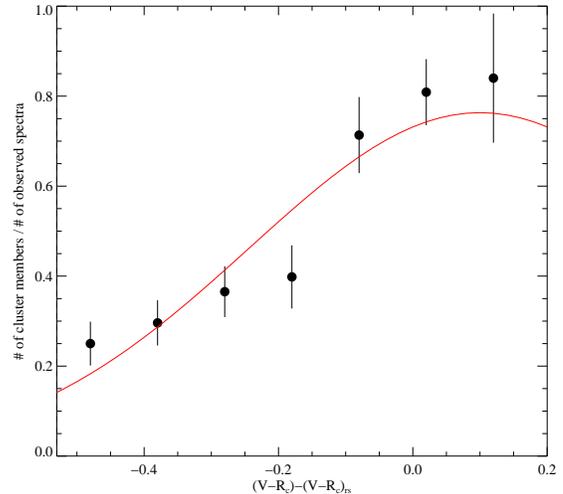}
\figcaption[f5.eps]{The probability of being a cluster member for
  all galaxies within our study region and brighter than our magnitude cut as
  estimated by the ratio between the number of cluster members and the number of
  galaxies with spectroscopic redshift. The x-axis is the relative V-R$_{c}$
  color with respect to the color of the red sequence, which is magnitude
  dependent (Fig.~\ref{cmd}). The fitted function (we use an ad-hoc Gaussian
  model) allows us to estimate the number of cluster members missed by our
  survey because of their extremely blue color. \label{membership}}
\end{figure}

Since it is impractical to cover the entire sample of galaxies spectroscopically
from the cluster core to the outskirt, we had to select targets in a way that
increases the efficiency of cluster member detections while minimizing any
potential bias. Since, at the time this project started, we did not have imaging
data in a sufficient number of bandpasses to determine credible photometric
redshifts, the target list for our spectroscopic observations is selected using
the color magnitude diagram for the V and R$_c$ filters (Fig.~\ref{cmd}), making
use of the cluster red sequence. To maximize completeness, we limit our survey
to relatively bright objects ($m_{\rm R_c}\leq 22.25$, which is, equivalently,
M$_{\rm Rc} \leq $M$^*$+2 at z=$0.55$) while adopting a generous color cut
toward the blue end of the distribution. The latter was adjusted in the course
of the survey such that its locus coincides with the V--R$_c$ color (relative to
the red sequence) at which the fraction of galaxies found to be cluster members
falls below 20\% (Fig. ~\ref{membership}). However, there will be unavoidably a
few extremely blue galaxies that are in fact cluster members but do not satisfy
our selection criteria. We estimate the resulting incompleteness by fitting a
heuristic Gaussian model to the data shown in Fig.~\ref{membership}.  If we
assume that the probability derived from the observed sample is representative,
and that the Gaussian fit can be extrapolated to yet bluer colors, we would
expect our survey to be incomplete at the $\sim 2\%$ level. We note, however,
that this incompleteness is severely color-dependent which, as we will show in
\S5.2, could potentially cause the fraction of emission-line galaxies, but not
the fraction of E+A galaxies, to be significantly underestimated.

\begin{figure}[h]
\epsscale{1.0} \plotone{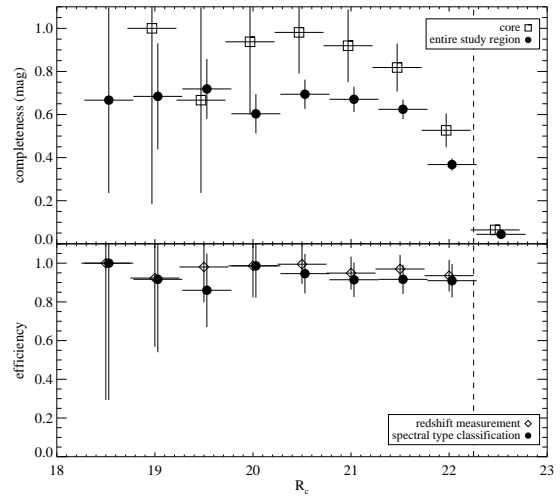}
\figcaption[f6.eps]{Completeness (top) and efficiency (bottom)
  of our spectroscopic survey as a function of galaxy magnitude. The color
  criterion of Fig.~\ref{cmd} has been applied. Top: survey completeness for the
  entire study region (filled symbols) and for Region A (open symbols). Bottom:
  survey efficiency, i.e. success rate for redshift determination (open symbols)
  and spectral classification (filled symbols). In both panels, the dashed line
  indicates the magnitude limit of the spectroscopic
  sample.  \label{completeness}}
\end{figure}

We define the completeness of our spectroscopic survey as the ratio of the
number of sources observed to the number of sources in the photometric catalogue
(within the color band marked in Fig.~\ref{cmd}), including spectra from which we
failed to measure a redshift\footnote{We include in this statistic a small
  number of stars which failed to be eliminated by the star-galaxy separation
  discussed in \S3.1}. In addition we define the efficiency of our survey as the
ratio of the number of high-quality spectra, from which both the redshift and
the equivalent width of spectral features can be measured accurately (again
within the color limits marked in Fig.~\ref{cmd}). The completeness and
efficiency as a function of galaxy magnitude are shown in the top panel of
Fig.~\ref{completeness}.  Inside the entire study region, the completeness is
roughly constant at about $70\%$ at R$_c< 21.75$, with the curve only dropping
within the very last bin before the magnitude limit at R$_c=22.25$. Also shown
in the top panel of Fig.~\ref{completeness} is the completeness within a radius
of 2 Mpc from the center of the cluster: again
the completeness is roughly independent of magnitude (except for the final bin
before our global magnitude limit) but now at a much higher level near unity.
Although the difference between these two curves shows that the completeness of
our survey is not spatially uniform, the relative spatial distribution of
galaxies of different types should be unaffected.

The bottom panel in Fig.~\ref{completeness} shows the efficiency in determining
the galaxy redshift and the efficiency for spectroscopic classification. The
former is almost perfect over the full magnitude range. The criteria used to
classify galaxies according to spectral type (defined in the following two
sections) are much stricter; still, the efficiency for spectroscopic
classification remains better than $0.8$ at all magnitudes, reflecting the good
signal-to-noise ratio of most of our spectra. Only a handful of spectra feature
very low signal-to-noise ratios which are caused either by poor slit positioning
or by erroneously high magnitudes reported by SExtractor. If the redshift can be
measured but no spectral type can be determined, the reasons are usually that
the key spectral features fall onto the chip gap or coincide with telluric
lines.

\subsection{Equivalent Width Measurement:}

In order to allow a comparison of our results with those from previous
spectroscopic studies of cluster galaxies we follow the approach of
\citet{tran03} to measure equivalent widths (EW), i.e.\ the inverse-variance
weighted integration of the flux over a spectral line, using integration windows
as defined in \citet{fish98}. EW of three spectral features are measured this
way: [OII] $\lambda\lambda 3727$, H$_{\delta}$, and H$_{\gamma}$. For those
galaxies with strong emission features but weak or non-detectable continuum, a
lower limit to the EW is calculated based on the variance of the spectrum near
the spectral feature in question. An exception is the case of the H$_{\beta}$
emission line for which the EW could be underestimated if absorption is present
too.  We account for this potential bias by subtracting the EW of the
H$_{\beta}$ absorption as provided by stellar population models \citep{bc03}
fitted to the data from 4060\AA\ to 5360\AA\ (rest frame) using the
template-fitting code \textit{pPXF} \citep{capp04}. We adopt the convention that
negative (positive) EW values indicate emission (absorption). Hereafter, we will
use the convention that $[OII]$ refers to the EW of [OII] in emission, whereas
$H_{\delta}$ and $H_{\gamma}$ refer to the EW of the respective lines in
absorption. Since we have fitted the emission and absorption components of
H$_{\beta}$ separately, the EW of H$_{\beta}$ in absorption and emission will be
specified as $H_{\beta,ab}$ and $H_{\beta,em}$.

\subsection{Spectral Type Classification:}

We classify cluster galaxies as emission-line galaxies, absorption-line
galaxies, and E+A galaxies, following the spirit of the classification used by
\citet{tran03} with a few modifications\footnote{Our goal is to classify
  galaxies according to activity, as reflected in their spectroscopic
  features. Unfortunately, a universally accepted system for such a
  classification has yet to be established (see the review in
  \citet{goto06}). This is especially true for E+A galaxies: since the exact
  locus of the dividing line between E+A galaxies and both emission- and
  absorption-line galaxies is ill-defined, the final classification in this grey
  area remains somewhat uncertain.}. The classification scheme of \citet{tran03}
only requires the relatively blue features [OII], H$_{\delta}$, and
H$_{\gamma}$, which is convenient for spectroscopic surveys of clusters out to
redshift $z\approx 1$ for which redder spectral lines are observationally
inaccessible in the optical passband. [OII] emission is an indicator of current
activity (be it star formation or the presence of an active nucleus), whereas
the two Balmer line features\footnote{It is worth pointing out in this context
  that historically the classification of E+A galaxies is based on the
  absorption features H$_{\delta}$ and H$_{\gamma}$, without any attempts at
  quantifying, or accounting for, emission components in these lines.} are
indicators of star formation within the last few hundred million years.  The
Balmer emission lines, however, are more robust diagnostics of star-formation
activity (e.g. \citet{mous06}) than [OII]. Since the latter is easily
underestimated because of extinction, the absence of [OII] emission does not
necessarily preclude the presence of star formation or other galaxy activity.
Although H$_{\alpha}$ is not accessible to optical surveys for galaxies at
$z>0.5$, the H$_{\beta}$ line of most of the galaxies in our target cluster
falls within the observable spectral range (although it comes dangerously close
to the broad atmospheric H$_2$O absorption band at 7600\AA). We therefore use
both $[OII]$ and $H_{\beta,em}$ to define emission-line galaxies, the specific
criteria being: $[OII] < -5 $\AA\ (the same cutoff as \citet{tran03}) or
$H_{\beta,em} < -5 $\AA. Galaxies with $[OII] > -5 $\AA\ and $H_{\beta,em} > -5
$\AA\ will be classified as absorption-line galaxies if $\frac{(H_{\delta} +
  H_{\gamma})}{2} < 4$\AA\ (the latter criterion being identical to the one used
by \citet{tran03}).

As for E+A galaxies, we deviate yet further from \citet{tran03} who simply
define E+A galaxies to be all galaxies that are neither emission- or
absorption-line galaxies. The definition adopted for this work is more
conservative in that we require an E+A galaxy to have {\em no} detectable
$[OII]$ or $H_{{\beta},em}$ and $\frac{( H_{\delta}+ H_{\gamma})}{2}>$6\AA\  (as
compared to the cutoff value of 4\AA\ of \citet{tran03}). This definition leaves
a small fraction of our galaxy sample unclassified, in recognition of the
aforementioned gray areas between E+A galaxies and both emission- and
absorption-line galaxies.

The motivation for our choice of more stringent criteria is twofold. As shown by
\citet{goto03} and \citet{goto07}, local samples of E+A galaxies compiled
without information about H$_{\alpha}$ may suffer from as much as $52\%$
contamination by H$_{\alpha}$ emission-line galaxies. \citet{blak04}, however,
found that a linear combination of the strengths of three Balmer absorption
lines (H$_{\beta}$, H$_{\delta}$, and H$_{\gamma}$) can be efficiently used to
identify this contaminating population. Following \citet{zabl96} we thus
calculate the average of $H_{\delta}$, $H_{\gamma}$, and $H_{\beta,ab}$ for all
galaxies in our sample that meet the more generous definition of E+A galaxies of
\citet{tran03}, and for which the H$_{\beta}$ line is observable.  We find
that only $45\%$ of them meet the criterion that $\frac{(H_{{\beta},ab}+ 
 H_{\gamma}+ H_{\delta})}{3}>5.5$\AA\ \citep{blak04}, implying that more
than half of the population thus selected may in fact feature H$_{\alpha}$ in
emission. If, however, we raise the threshold value for $\frac{(
H_{\gamma}+H_{\delta})}{2}$ from 4 to 6\AA, then $100\%$ of the galaxies
thus selected, and for which the H$_{\beta}$ line is available, fulfill the
criteria of \citet{blak04}. The second motivation for our more stringent
definition of E+A galaxies is that the detection of [OII] in a galaxy is not an
unambiguous sign of star formation but could also be indicative of non-stellar
radiation. Indeed, \citet{yan06} find that a large fraction of post-starburst
galaxies may be misidentified as undergoing current star formation, although the
[OII] emission actually originates from AGN. Since our project has currently no
means of efficiently distinguishing between star-bursting galaxies and
post-starburst galaxies containing AGN, and since the compilation of a
statistically complete sample of post-starburst galaxies is not the main goal of
this work, we exclude from our E+A sample all galaxies featuring {\em any}\/
[OII] emission, in order to keep the resulting sample as clean as possible and
avoid complications caused by the presence of AGN for the interpretation of the
mechanisms responsible for the termination of star formation activity.  Although
the sample of E+A galaxies selected by our more conservative criteria thus
misses a possibly sizeable and interesting fraction of the entire post-starburst
population as suggested in \citet{yan06}, it is better suited to study the
interactions between galaxies and the intra-cluster medium, and it is also less
contaminated by weakly star-forming galaxies.

Following \citet{dres99}, we can classify the emission-line galaxies further
into e(a), e(b), and e(c) sub-types based on the strength of the $H_{\delta}$
absorption and $[OII]$ emission lines. Both of these features are indicative of
starbursts within a few million years to one billion years before the time of
observation. Specifically, e(a) galaxies are defined as systems with $H_{\delta}
> 4$\AA\ and $-40$\AA$<[OII]<-5$\AA, whereas e(b) galaxies feature
$[OII]<-40$\AA, regardless of $H_{\delta}$; the spectral evidence of young and
massive stars in either of these subtypes suggests recent starbursts. All other
emission-line galaxies, i.e.\ systems lacking signs of explosive star formation
in the recent past are classified as e(c). The difference between e(b), e(a),
and E+A galaxies is thus the decreasing strength of the [OII] emission line.
The absence of [OII] in E+A galaxies suggests that the star formation activity
is halted, whereas, at the opposite extreme, strong [OII] emission makes the
e(b) sub-type promising candidates of starburst galaxies. The, in terms of [OII]
strength, intermediate e(a) galaxies, on the other hand, are not easily placed
in this sequence since, as mentioned before, the strength of [OII] emission is
highly sensitive to dust extinction. \citet{pogg00}, for instance, find about
half of a sample of very luminous infrared galaxies
[log(L$_{IR}$/L$_{\sun}$)$>$11.5] to exhibit e(a) spectra. The distinction
between these three populations is thus ambiguous; some of the E+A galaxies may
actually be e(a) galaxies whose [OII] emission is completely extinguished by
dust. Conversely, some of the e(a) galaxies could in fact be post-starburst
galaxies with some residual star formation. In spite of these ambiguities, we
emphasize that e(a) and E+A galaxies are in general distinct populations, which
is -- as we shall demonstrate -- reflected in their different distribution
within the cluster environment. The relevance of e(a) and e(b) galaxies in the
context of this study is the potential for both galaxy types to trace the E+A
progenitor population.

For clarity, the definition of the spectral types is summarized in
Table~\ref{spectdef}. The number of galaxies of each type is given in
Table~\ref{spectsum}.  Note that, because of our conservative definition of E+A
galaxies, about 8\% of all cluster members with spectroscopic information are
missing in Table~\ref{spectsum} as well as in the analysis that follows. They
are the population with weak [OII], weak H$_{\beta}$, and Balmer absorption
lines whose strength falls between the absorption-line and E+A criteria
(4\AA$<\frac{(H_{\delta}+H_{\gamma})}{2}<$6\AA). They are probably a mixture of
dusty star-forming galaxies and post-star-forming galaxies. In \citet{tran03}
and \citet{tran04}, galaxies of this type are also classified as E+A galaxies.

\begin{deluxetable*}{cl}
\tabletypesize{\scriptsize}
\tablewidth{0pc}
tablecolumns{2} 
\tablecaption{Definition of spectral types\label{spectdef}}
\tablehead{ 
\colhead{Type} & \colhead{Criteria} }  
\startdata
Emission-line     & $[OII] < -5 $\AA\ or $H_{\beta,em} < -5 $\AA\  \\
Absorption-line  & $[OII] > -5$\AA\, $H_{\beta,em} > -5 $\AA\, and $\frac{(H_{\delta} +H_{\gamma})}{2} < 4$\AA\ \\
E+A                      & no detection of $[OII]$ and $H_{\beta,em}$, and $\frac{(H_{\delta} + H_{\gamma})}{2} > 6$\AA\ \\
\cline{1-2}\vspace*{-2mm} \\
e(a)             & Emission-line galaxies with $H_{\delta}> 4$\AA\, and $-40$\AA$<[OII]<-5$\AA \\
e(b)             & Emission-line galaxies with $[OII]<-40$\AA \\
e(c)             & Emission-line galaxies with $H_{\delta} < 4$\AA\, and $-40$\AA$<[OII]<-5$\AA
\enddata
\end{deluxetable*}

\begin{deluxetable}{ccc}
\tabletypesize{\scriptsize}
\tablewidth{0pc}
\tablecolumns{3} 
\tablecaption{Spectral types summary\label{spectsum}}
\tablehead{ 
\colhead{Type} & \colhead{All} & \colhead{cluster members} }  
\startdata

all spectra\tablenotemark{a}         &  1368   & \nodata         \\
sample spectra      &  1147   & \nodata \\
redshift measurements   &  1103   &     563         \\
spectra with spectral diagnostics & 1023  &   530    \\
\cline{1-3} \vspace*{-2mm} \\
Emission-line galaxies   &  507   &  203            \\
Absorption-line galaxies &  426  &  260            \\
E+A galaxies             &  21    &   17            \\     
Unclassified             &  69    &    50            \\
\cline{1-3} \vspace*{-2mm} \\
e(a) & 76 & 42 \\
e(b) & 66 & 27 \\
e(c) & 365& 134 
\enddata
\tablenotetext{a}{No magnitude or color cuts are applied. Spectra obtained serendipitously are included.}
\end{deluxetable}

\section{Result:}

\subsection{Photometric Redshift}
\begin{figure}[h]
\epsscale{0.8} 
\plotone{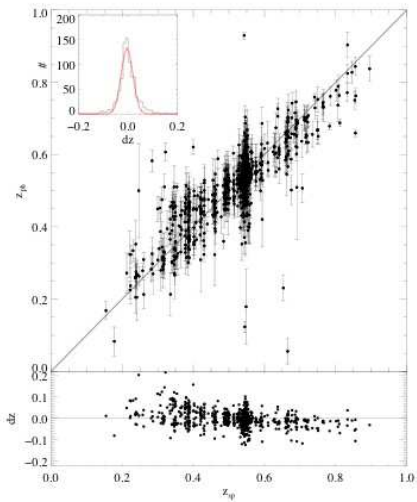}
\figcaption[f8.eps]{Photometric versus spectroscopic redshifts for
  galaxies in the field of MACSJ0717.5+3745; we only show galaxies from our
  \textit{Le Phare} training set (see text for details).  The lower panel shows
  the residuals d$z$ as a function of the spectroscopic redshift, while the
  insert in the top left corner shows the histogram of the residuals as well as
  a Gaussian model used to estimate their standard deviation $\Delta
  z$. [The full resolution figure will be published in ApJ]\label{pzVSspz}}
\end{figure}

From our photometry in the u*, B, V, R$_c$, I$_c$, and z$^\prime$ bands,
photometric redshifts are derived for galaxies with R$_c<$24.0 ($\sim$~M$^*$+4
at z$=$0.55) using the adaptive SED-fitting code \textit{Le Phare}
\citep{arno99,ilbe06}. \textit{Le Phare} employs ${\chi}^2$ optimization,
comparing the observed magnitudes with those predicted from an SED library. An
adaptive method is applied to adjust the photometric zero-points for all
passbands by using the sample of spectroscopic redshifts as a training
set\footnote{The resulting correction of typically 0.05 mag is consistent with
  the statistical and systematic uncertainty of our photometric
  calibration.}. No Bayesian prior for the galaxy redshift distribution is
assumed. We use the library of empirical templates of \citet{ilbe06} and the
Calzetti extinction law \citep{calz01} for reddening corrections applied to
templates later than Sbc. The training set is obtained from the spectroscopic
redshift database by selecting all galaxies with an R$_c$-band magnitude
brighter than 21.5 and a minimal separation from the closest SExtractor-detected
neighbor of $3.5 \arcsec$ to avoid photometric errors from close pairs or
blended sources. The resulting training set comprises 273 galaxies at redshifts
ranging from 0.15 to 0.9. The comparison of spectroscopic and photometric
redshifts is shown in Fig.~\ref{pzVSspz}. The statistical error of the
photometric redshifts, as derived from the Gaussian distribution of the
residuals shown in the insert of Fig.~\ref{pzVSspz}, is found to be $\Delta z =
0.024$.

The main purpose of the photometric redshift catalogue is to allow us to
estimate the local projected galaxy density with better statistics than would be
feasible from the spectroscopic data set alone. On the basis of photometric
redshifts, we define cluster members to be the galaxies with $| z_{ph} - z_{cl}
| < \sigma_{ph-z}$, where $\sigma_{ph-z} = (1+z_{cl}) \Delta z$. Note that
this redshift cut ($\pm \sigma_{ph-z} = \pm0.036$) is much more generous than
that used before for the spectroscopic redshifts ($\pm3\sigma = \pm 0.022$); the
chosen cutoff value represents a good compromise between completeness and
contamination. The resulting catalogue of ``photometric'' cluster members
includes 2218 galaxies and allows us to estimate the local galaxy density in the
entire Suprime-Cam field of view.  The much higher accuracy of our spectroscopic
redshifts, on the other hand, provide us with an opportunity to study the
dynamical structure of the cluster-filament system in detail.

\begin{figure}[h]
\plotone{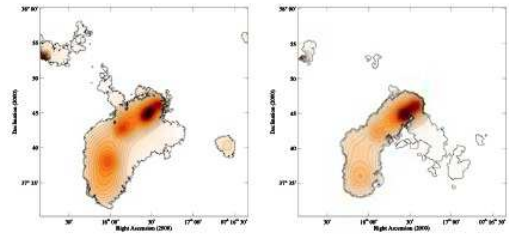}
\figcaption[f7a.eps]{Surface density of cluster members using
  either photometric redshifts (top) or selection by red-sequence color (bottom)
  \citep{jeyh08}. The contour levels are given by $1.2^n \Sigma_{bkg}$, where
  $\Sigma_{bkg}$ is the background surface density of each plot. The bold,
  ragged contour marks the boundary of regions outside of which the significance
  of any structure in these adaptively smoothed images falls below the 3$\sigma$
  confidence level. [The full resolution figure will be published in ApJ] \label{galdenmap}}
\end{figure}

Using the photometrically selected set of cluster members, we map the surface
density of cluster galaxies. A smoothed version of this distribution (shown in
Fig.~\ref{plain}) is obtained with the adaptive-smoothing algorithm
\textit{asmooth} \citep{ebel06} which preserves the signal-to-noise ratio of all
significant features on all scales. The comparison of the result with the maps
derived by \citet{ebel04} and \citet{jeyh08} using only galaxies on the cluster
red sequence proves instructive in as much as it highlights the dominance of
blue galaxies in low-density, filamentary environments. We show in
Fig.~\ref{galdenmap}, side by side, the galaxy surface density maps obtained by
us using photometric redshifts as described above, and the equivalent map from
\citet{jeyh08} which is based entirely on galaxies with V--R$_c$ colors
consistent with the cluster red sequence. Note the dramatic difference in the
relative prominence of the main cluster and the filament. 

In order to allow a direct comparison with the results from earlier work on the
impact of cluster environment on galaxy activity \citep{koda01, balogh04, sato06},
we also calculate for all ``photometric'' cluster members the projected local
density $\Sigma_{10}$, defined as the galaxy number density within a circle
whose radius is given by the separation to the $10^{th}$ closest neighbor. The
resulting values of $\Sigma_{10}$ are consistent with the surface density
distribution shown in the top panel of Fig.~\ref{galdenmap}.

\subsection{Broad-Band Colors of Cluster Members of Different Spectral Types}
\begin{figure}[h]
  \epsscale{1.0} \plotone{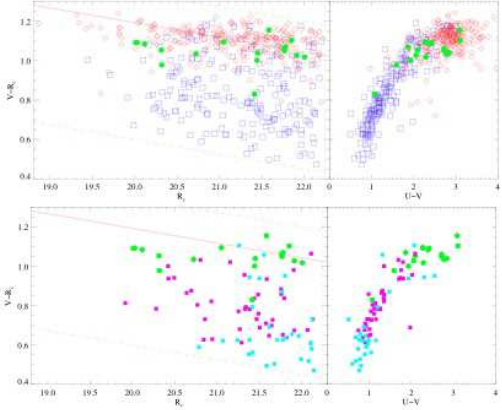}  
  \figcaption[f9.eps]{Top: (left) Color-magnitude diagram for galaxies of
    different spectral type. Open red diamonds mark absorption-line galaxies,
    open blue squares emission-line galaxies, and filled green circles E+A
    galaxies. The dashed lines are the color and magnitude cuts from
    Fig.~\ref{cmd}. (right) U-V v.s. V-R$_c$ of galaxies of different spectral
    type. Bottom: same as top panel, but showing E+A galaxies in comparison with
    their hypothesized progenitor population, e(a) and e(b) galaxies (magenta
    and cyan symbols respectively). [The full resolution figure will be published in ApJ] \label{cc}}
\end{figure}

The comparison of Fig.~\ref{cmd} with the color-magnitude diagram for different
spectral types of galaxies (left panel of Fig.~\ref{cc}) illustrates the
potential bias caused by the blue color cut discussed in \S4.1. Not
surprisingly, all of the bluest cluster members found in our study are
emission-line galaxies, the fraction of which is thus likely to be
underestimated. This is particularly true for the e(b) subtype which we find to be
heavily concentrated at the blue, faint end of our selection in color-magnitude
space. If all cluster members bluer than our color cut are in fact e(b) galaxies
-- clearly the worst-case scenario --, the number of e(b) galaxies in
MACSJ0717.5+3745 could be underestimated by as much as $\sim$ 30\%. We shall keep
this in mind when, later, discussing the fraction of e(b), e(a), and
emission-line galaxies in general. Note, however, that the E+A and
absorption-line galaxies lie well away from the color cut, so that their numbers
can be expected to be unaffected by color-dependent incompleteness.

Figure~\ref{cc} compares the magnitude and color distributions of cluster
members of different spectral type. As has been noted before by other authors,
for example \citet{dres83}, \citet{tran03}, and \citet{tran07}, E+A galaxies are
found to exhibit broad-band colors that are intermediate between those of
absorption- and emission-line galaxies, as is to be expected from their spectral
definition as a
composite of old and new stellar populations. Differences between the various
spectroscopic types can be characterized by measuring the mean and scatter of
the relative V-R$_{c}$ color of a given galaxy class with respect to the red
sequence (Table~\ref{colorstats}). The mean relative color of zero found for
absorption-line galaxies confirms our linear fit to the cluster red sequence
calculated without selection of any specific spectroscopic types
(Fig.~\ref{cmd}). The small scatter in the relative V-R$_{c}$ color of
absorption-line galaxies is consistent with the result of, e.g.,
\citet{blakeslee06} and \citet{tran07} that early-type cluster members are
formed at $z\sim 2$. The mean relative colors of the other spectroscopically
defined galaxy types discussed in this study are, in order of increasing
blueward distance from the red sequence, E+A galaxies, e(a), and e(b) galaxies,
a ranking consistent with the hypothesis that E+A galaxies represent an
intermediate stage in the evolutionary sequence from active emission-line
galaxies to passively evolving early-type galaxies. Interestingly, this scenario
is also supported by the different scatter around the mean relative color for
the different populations: the color scatter of the E+A galaxies is
significantly smaller than the one of e(a) and e(b) galaxies, their potential
precursors \citep{pogg99}, but still larger than the one of the evolved
population of absorption-line galaxies forming the cluster red sequence.
Although their spectral properties imply that E+A galaxies evolved from the e(a)
and/or e(b) population less than one billion years ago, their color
characteristics have thus already become remarkably similar to those of
absorption-line galaxies.

Another interesting fact that emerges from Fig.~\ref{cc} relates to the mean
color of e(b) galaxies. Their locus at the extreme blue end of the distribution
shown in the left bottom panel of Fig.~\ref{cc} is not solely the result of the
strong [OII] emission to which they owe their definition. Since [OII] falls into
the V band at $z=0.55$, the fact that the majority of the e(b) galaxies found in
our study also exhibits very blue U-V colors (right bottom panel of
Fig.~\ref{cc}) indicates significant continuum emission in the UV rest frame.
Follow-up observations are needed of the few e(b) galaxies that are found to be
unusually red to establish whether nuclear activity is responsible for the
strong [OII] emission observed.

\begin{deluxetable}{lcc}
\tabletypesize{\scriptsize}
\tablewidth{0pc}
\tablecolumns{3} 
\tablecaption{Relative color statistics of spectroscopic types\label{colorstats}}
\tablehead{
\colhead{}    & \colhead{mean\tablenotemark{a}}& \colhead{scale\tablenotemark{a}}                 
}
\startdata

Absorption-line galaxies & $0.00^{+0.005}_{-0.004}$ &  $0.0068^{+0.0004}_{-0.0004}$  \\
E+A galaxies                     & $-0.05^{+0.03}_{-0.03} $   &  $0.08^{+0.01}_{-0.03} $\\
Emission-line galaxies\tablenotemark{b}    & $-0.26^{+0.01}_{-0.02} $   &  $0.176^{+0.009}_{-0.006} $\\
\cline{1-3}
e(a) galaxies\tablenotemark{b}                     & $-0.30^{+0.02}_{-0.03}$     &  $0.12^{+0.01}_{-0.02}$ \\
e(b) galaxies\tablenotemark{b}                     & $-0.42^{+0.04}_{-0.04}$     & $0.18^{+0.02}_{-0.06}$ \\
e(c) galaxies\tablenotemark{b}                     & $-0.23^{+0.02}_{-0.02}$     & $0.175^{+0.01}_{-0.009}$ 
\enddata
\tablenotetext{a}{Mean and scale are calculated using the biweight estimator in \citet{beers90}}
\tablenotetext{b}{Because of the blue color cut mentioned in \S4.1, the mean and scattering of the emission-line galaxies and the subtypes, in particular the e(b) galaxies, may be biased.} 
\end{deluxetable}

\subsection{Dynamical Scales}

\begin{deluxetable}{lcl}
\tabletypesize{\scriptsize}
\tablewidth{0pc}
\tablecolumns{3} 
\tablecaption{Cluster property\label{xrayprop}}
\tablehead{
\colhead{}    & \colhead{value}& \colhead{unit}                 
}
\startdata

$f_x$\tablenotemark{a}         & $2.74\pm 0.03$ & $10^{-12}$~ergs~s$^{-1}$~cm$^{-2}$\\
$L_x$\tablenotemark{a}        & $24.6\pm 0.3 $ & $10^{44}$~ergs~s$^{-1} $  \\
$kT$\tablenotemark{a}          & $11.6\pm 0.5$  & kev           \\
$\beta$                                      & $1.1\pm 0.1$   &                                 \\
$\rho_{0}$                                 & $1.71\pm 0.05$ & $10^{14}~M_{\sun}$~Mpc$^{-3}$     \\
$r_c$                                          & $92 \pm 6 $    & arcsec                        \\
\cline{1-3}
z$_{cl}$\tablenotemark{b}                        & $0.5446\pm 0.0005$ & \\
$\sigma$\tablenotemark{b}                      & $1612 \pm 70$ & km~s$^{-1}$ \\      
$R_{\rm vir,gal}$\tablenotemark{c}        & $3.0\pm 0.2$           & Mpc    \\
$R_{\rm vir,x-ray}$ \tablenotemark{c}    & $2.9\pm 0.1$     & Mpc \\
$M_{x}$\tablenotemark{c}                        & $28\pm 5$        & $10^{14}$~M$_{\sun}$ \\
$R_{rs}$                       & $1.9$      & Mpc                  
\enddata
\tablenotetext{a}{Ebeling et al. (2007)} \tablenotetext{b}{The values for the
  mean redshift z$_{cl}$ and velocity dispersion $\sigma$ are calculated
  iteratively using the redshifts of cluster members within the virial radius (2.9
  Mpc) and thus differ slightly from the values published by \citet{ebel07}
  where a radius of 1 Mpc was used.}  \tablenotetext{c}{The uncertainty is
  estimated by error propagation.}
\end{deluxetable}

A fundamental problem of many forms of dynamical analysis is the inherent
assumption of virial equilibrium. Models appropriate for the description of
clusters in the process of active assembly, such as MACSJ0717.5+3745 or --
albeit to a lesser degree -- MS0451.6-0305 \citep{mora07}, would have to take
into account the effects of mergers and complex substructure on a wide range of
scales. With the exception of the simplest cases this is infeasible, in part
owing to our insufficient knowledge of the true three-dimensional geometry of the
system. With this caveat in mind, we note that the simplistic dynamical scales
derived in this section should only be considered crude estimates.

The most relevant dynamical scales for this work are the virial radius $R_{vir}$
and the ram-pressure stripping radius $R_{rs}$. To obtain estimates of these
quantities, we need a description of the cluster's gas density profile as well
as a global X-ray temperature. The former is obtained by fitting a $\beta$-model
\citep{cava76} 
\begin{equation}
\rho_{gas} =  \rho_{0}\left[1+\left(\frac{r}{r_c}\right)^2\right]^{-\frac{3}{2}\beta} \label{beta}
\end{equation}
to the observed X-ray surface brightness. Specifically, the center of the X-ray
emission is determined by fitting a two-dimensional elliptical $\beta$-model to
the exposure-weighted image after point sources have been removed. The radial
profile is then extracted using this center position.  The subsequent fit of a
one-dimensional $\beta$-model is limited to radii larger than $11\arcsec$ (70
kpc) to minimize the impact of the complex merging features in the heavily
disturbed cluster core. A global X-ray temperature is measured by extracting the
X-ray spectrum from $r=$70 kpc to $r=$0.94 Mpc (r$_{1000}$) and using
\textit{Sherpa} to fit a MEKAL plasma model \citep{mewe85} with the absorption term
frozen at the Galactic value. From the X-ray temperature and the $\beta$-model
profile, we estimate the virial radius $R_{vir}$ using the formula of
\citet{arna02}
\begin{eqnarray*}
R_{vir} & = & 3.80~\beta_T^{1/2}\Delta_z^{-1/2}~(1+z)^{-3/2} \nonumber \\
        &  & \times(\frac{kT}{10{\rm keV}})^{1/2}h_{50}^{-1}~{\rm Mpc} \nonumber
\end{eqnarray*} 
with
\begin{eqnarray*}
\Delta_z  =  (200\Omega_0)/(18\pi^2\Omega_z). \nonumber
\end{eqnarray*}
Here $\beta_T$ is the normalization of the virial relation, i.e.,
$GM_v/(2R_{vir})= \beta_T~{\rm k}T$.  The virial radius thus derived is about
2.9 Mpc, which is consistent with the virial radius (3 Mpc) estimated from the velocity
dispersion using the relation \citep{gunn72}
\begin{eqnarray*}
R_{vir} = 1.7{\rm h}^{-1}{\rm Mpc} \frac{\sigma}{1000\,{\rm km~s}^{-1}} [\Omega_m(1+z)^{3}+\Omega_{\Lambda}]^{-0.5}.
\end{eqnarray*}
The ram-pressure stripping radius is estimated by equating the gas pressure
required to strip a Milky-Way-like galaxy from all its gas to the gas density
obtained previously in our X-ray analysis (Eqn.~\ref{beta}). The former can be
obtained from the stripping requirement \citep{gunn72, treu03}
\begin{eqnarray*}
\rho_{gas}v_i^2  & > &  2.1\cdot 10^{-12} {\rm N m}^{-2} \left(\frac{v_{rot}}{220\, {\rm km~s}^{-1}}\right)^2\left(\frac{r_h}{10\,{\rm kpc}}\right)^{-1} \nonumber\\
                &   & \times \left(\frac{\Sigma_{HI}}{8\cdot10^{20} m_H\,{\rm cm}^{-2}}\right)
\end{eqnarray*} 
Here $v_i$ is the velocity of the infalling galaxy which, in all other respects
(rotation velocity $v_{rot}$, scale length $r_h$, and HI surface density
$\Sigma_{HI}$), is assumed to be similar to the Milky Way, as indicated by the
units used above. The infall velocity $v_i$ is estimated as
\begin{eqnarray*}
v_{\rm i} (r)= \sqrt{\frac{2GM_{\rm x}}{r}-\frac{GM_{\rm x}}{R_{vir}}},
\end{eqnarray*}
where $M_{\rm x}$ is the total mass within $R_{vir}$ derived from the X-ray
properties of the cluster (see, e.g., \citet{maug03}. A summary of the results
as well as of related X-ray cluster properties is given in Table~\ref{xrayprop}.

Additional dynamical scales, that are of less interest to us here, can be
computed to assess the efficiency of tidally triggered star formation, tidal
stripping, starvation, harassment, and galaxy mergers. As discussed by
\citet{treu03}, the first two effects are only important within a few hundred
kpc of the cluster core and are thus difficult to study statistically for any
single cluster. In addition, the complex X-ray morphology in the core of
MACSJ0717.5+3745 makes this system particularly ill-suited for quantitative
modeling. Starvation and harassment, on the other hand, are efficient at larger
radii, with the starvation radius being equivalent to the virial radius \citep{balogh00}, and
harassment being almost independent of cluster-centric distance \citep{moor96,moor98}.  Since
probably all cluster galaxies under study here are thus subject to harassment,
it would not be easy to isolate the impact of this particular effect. The last
mechanism, galaxy merging, is largely irrelevant in the densest parts of the
cluster environment, owing to the large velocity dispersion, but peaks near the
virial radius (see e.g. \citep{ghigna98}).

\subsection{Redshift Distribution of Cluster Members}

\begin{figure}[ht]
\plotone{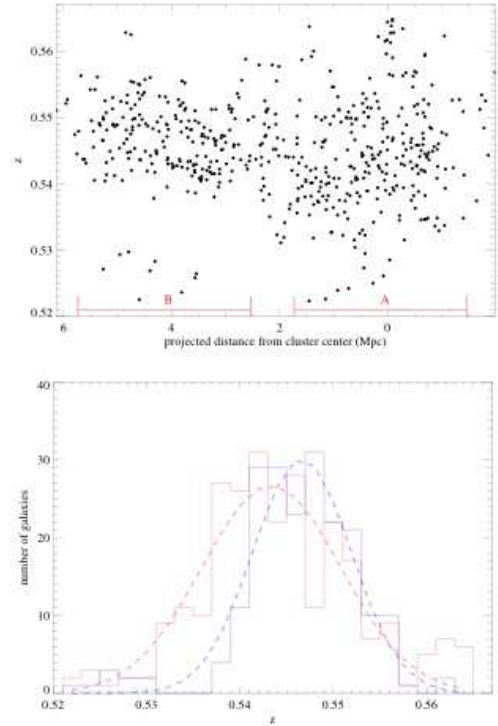}
 \figcaption[paper_zonaline2.ps]{
  Top: redshift distribution of cluster members as a function of their projected
  distance from the cluster center, measured along the line shown in
  Fig.~\ref{plain}. Bottom: redshift histogram for galaxies in the two regions A
  and B of Fig.~\ref{plain}; the red lines refer to Region A (cluster center),
  the blue lines to Region B (filament). [The full resolution figure will be published in ApJ] \label{zonaline}}
\end{figure}

The dynamical scales derived in the previous section refer to the primary
cluster, but exclude the filament. As a first look into the properties of the
large-scale structure observed, all the way from the cluster core to the
extended peak in the galaxy density in Region B, we show in Fig.~\ref{zonaline}
the distribution of galaxy redshifts projected onto a line starting northwest of
the cluster center and ending southeast of the filament, as indicated in
Fig.~\ref{plain}. We find the average redshift of the galaxies in Region B,
$\bar{z}_B = 0.5468\pm0.0004$, to be significantly higher than that of the
galaxies around the main cluster, $\bar{z}_A = 0.544\pm0.0006$. The velocity
difference between these two regions is thus 630 km/s. In addition, the
velocity dispersion of the galaxies in Region B is $\sigma_{z,B} = (1050\pm65)$
km/s, significantly less than the value of $\sigma_{z,A} = (1630^{+75}_{-93})$ km/s
found in Region A. Although the Chandra ACIS-I image of MACSJ0717.5+3745 covers
a fair fraction of Region B, it just misses the galaxy density peak at its
center. Since, in addition, the exposure of this observation is only 60 ks, we
are currently unable to obtain an estimate of the gas temperature and mass in
Region B. Deriving a mass estimate from $\sigma_{z,B}$ is not meaningful because
of the clearly unvirialized nature of this structure.

\subsection{Spatial Distribution of Cluster Members}

\begin{figure}[h]
\epsscale{1.0} \plotone{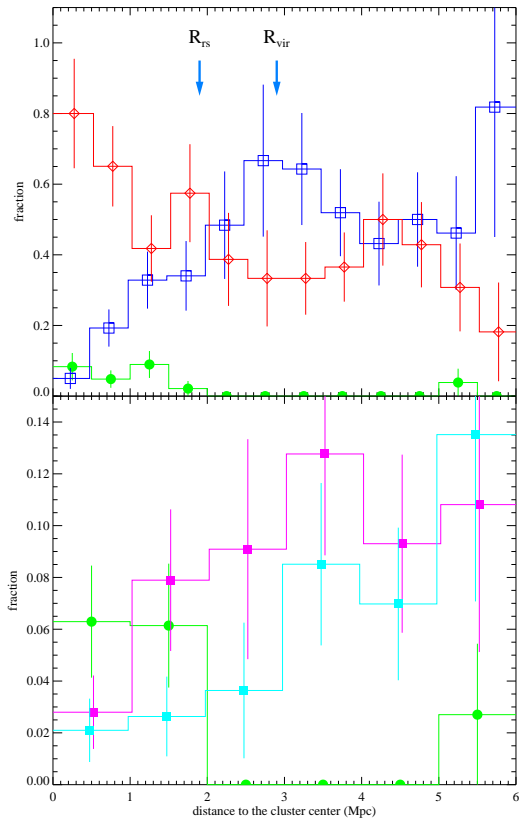}
\figcaption[f11.eps]{Distribution of galaxies of different
  spectral type with cluster-centric distance. The symbols the same as in
  Fig.~\ref{cc}. In each bin the fraction is calculated by dividing the
  number of member galaxies of a given spectral type by the total number of
  member galaxies observed spectroscopically. The plotted error bars assume
  Poisson statistics. Some symbols are shifted slightly by $\pm0.02$ Mpc for
  clarity. \label{galxradial}}
\end{figure}

We show in Fig.~\ref{galxradial} the distribution of different galaxy types as a
function of (radial) distance to the cluster center. The fraction of
absorption-line galaxies drops from $\sim$80\% in the cluster core to $\sim$30\%
at the virial radius, where the fraction of emission-line galaxies shows the
opposite behavior, increasing from about 5\% at the cluster center to almost
70\% at the virial radius. Beyond 3.5 Mpc, however, both trends are reversed as
we approach the galaxy density peak of Region B at 4.5 Mpc. At the largest radii
probed by our study, the ratio of the numbers of absorption- and emission-line
galaxies begins to approach the field value. An interpretation of our data in
the context of the well known morphology-density relation\citep{dres80} will be
given in the next section.

Interestingly, the distribution of E+A galaxies follows neither of the above
trends.  Rather, almost all E+A galaxies (16 out of 17) are found to reside in
Region A. No E+A galaxy is detected at cluster-centric distances exceeding 2
Mpc, with the sole exception of a single specimen near the southern edge of
Region B. Remarkably, the radial locus of the abrupt decline in the number of
E+A galaxies thus coincides almost perfectly with the ram-pressure stripping
radius, $R_{rs}=1.9$ Mpc. 

The spatial distribution of starburst galaxies is very different again: the
distribution of both e(a) and e(b) galaxies follows, qualitatively, the
distribution of their parent type, emission-line galaxies in general. Comparing
the e(a) and e(b) distributions, the fraction of e(b) galaxies appears to show a
more gradual and monotonic rise with cluster-centric distance than that of e(a)
galaxies. The difference between the e(a) and e(b) distributions is, however,
not significant: a Kolmogorov-Smirnov test finds the probability of them being
drawn from the same parent distribution to be 0.69. We note though, that the
aforementioned incompleteness of up to 30\% in our e(b) sample would likely
increase the fraction of e(b) galaxies at all cluster-centric radii.

\begin{figure*}[h]

\epsscale{0.48} 
\plotone{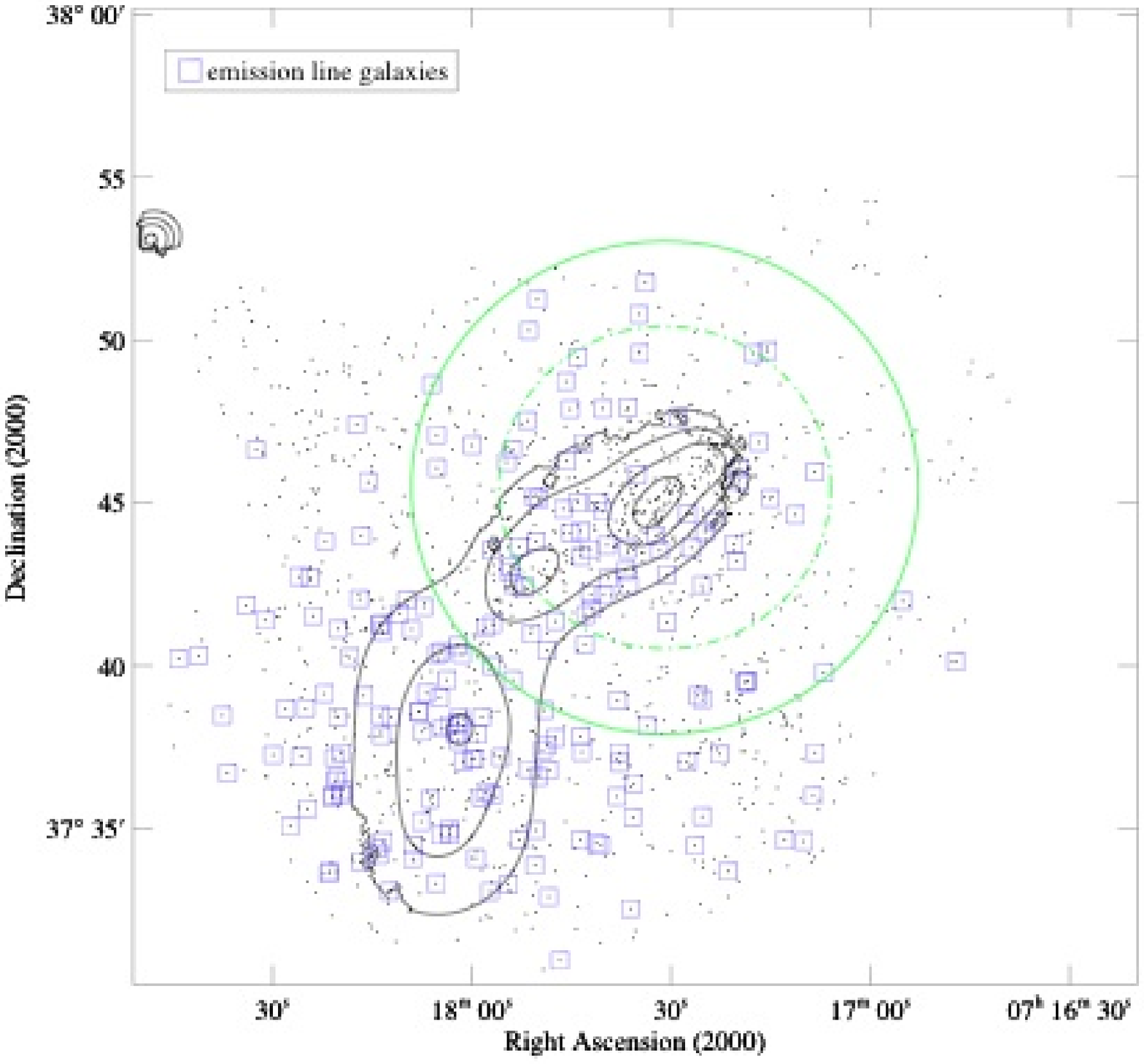}
\plotone{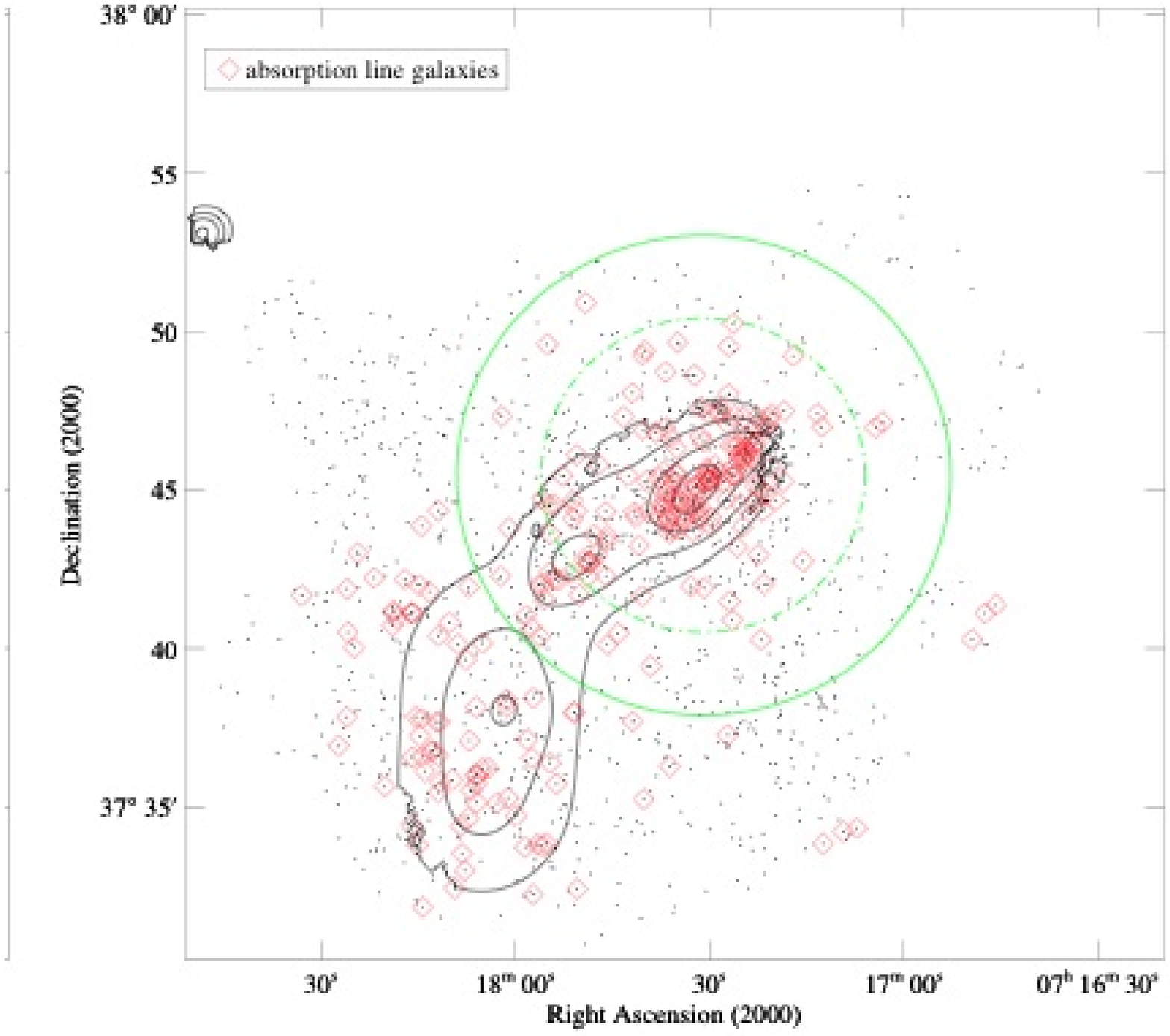}
\plotone{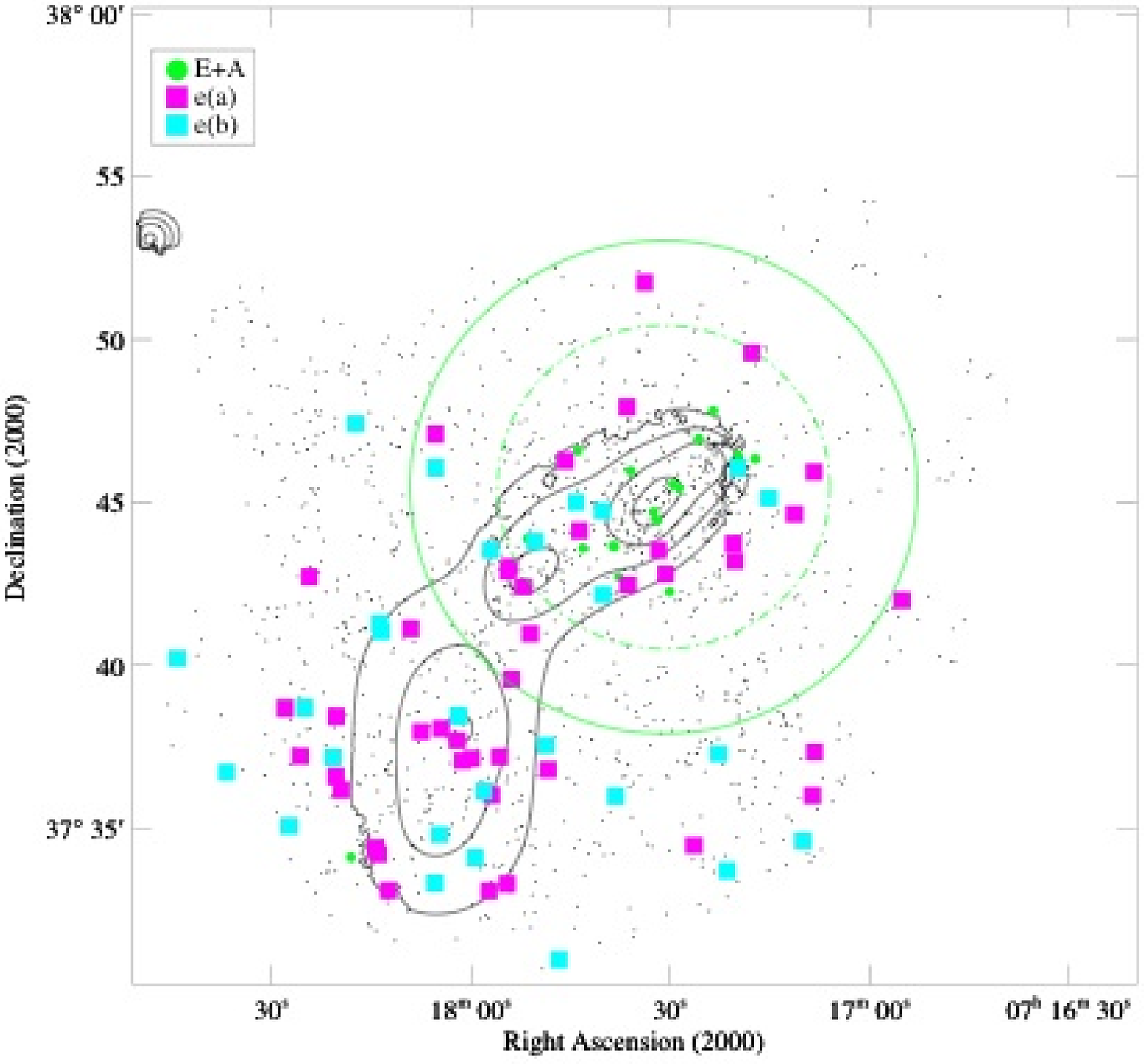}

\figcaption[paper_galxtype_distr1.ps]{Projected spatial distribution of
  different galaxy types in the MACSJ0717.5+3745 system. Symbols are as in
  Fig.~\ref{cc}, with the addition of small black dots which mark the locations
  of all galaxies within our study region that have been observed
  spectroscopically. Also shown are the galaxy-density contours from
  Fig.~\ref{plain}. 
The green circles have radii of $R_{vir}$ (dashed) and $R_{rs}$ (solid),
  respectively, as estimated in \S3.3 [The full resolution figure will be published in ApJ]\label{galxdistr}}
\end{figure*}

Although more difficult to interpret quantitatively, the two-dimensional spatial
distribution of galaxies of different spectral types provides additional
insight. To facilite a visual analysis, we show, in Fig.~\ref{galxdistr}, the
respective distributions separately for emission-line galaxies, absorption-line
galaxies, and -- in a third panel -- E+A, e(a), and e(b) galaxies. The contours
of the projected galaxy density of Fig.~\ref{galdenmap} (top panel) are
overlaid. We compare the distributions shown in these three panels in the
following.

The absorption-line galaxies are strongly concentrated in the double core of
MACSJ0717.5+3745 proper, a region which is almost entirely devoid of
emission-line galaxies. High concentrations of absorption-line galaxies are also
found at the locations of two galaxy groups embedded in the filament, both of
which are also detected in X-rays (Fig.~\ref{xrayim}). The string of three
high-density regions near the nominal cluster center thus causes an excess of
absorption-line galaxies in Region A, which extends linearly over almost 3 Mpc
northwest to southeast of the cluster core. By contrast, the observed
overdensity of cluster members in Region B has a much less pronounced effect on
the distribution of absorption- and emission-line galaxies: overall, both galaxy
types exhibit similar concentrations. Interestingly, absorption-line galaxies in
this region are, however, found preferentially south of the center of Region B,
whereas emission-line galaxies dominate the northern part of this area, roughly
0.5 Mpc away in projection. It is this large-scale segregation between
absorption- and emission-line galaxies that is responsible for the shift of the
overall distribution of cluster members in Region B shown in the top and center
panel of Fig.~\ref{galdenmap}; the apparent, not well defined center of this
region is inhabited by a concentration of e(a) and e(b) galaxies (bottom panel
of Fig.~\ref{galdenmap}).  Relating these observations to the true
three-dimensional distribution of galaxy types in Region B is, however, greatly
complicated by both projection effects and the likely presence of significant
peculiar velocities \citep{barrett05}. Looking at yet larger scales, it is clear
from Fig.~\ref{galxdistr} that the distribution of emission-line galaxies is
vastly more extended over our study region than is the distribution of
absorption-line galaxies. The lack of pronounced overdensities in the
distribution of emission-line galaxies suggests strongly that a significant
fraction of emission-line galaxies observed apparently near high-density regions
of the MACSJ0717.5+3745 system are in fact projected there from much larger
distances along the line of sight. This point is worth keeping in mind for the
interpretation of the true spatial distribution of all types of emission-line
galaxies.

E+A galaxies, finally, are found to be distributed almost evenly within the
ram-pressure stripping radius $R_{rs}$; we again note, however, the likely
presence of projection effects. Since the small number of E+A galaxies in our
sample does not allow us to derive a deprojected radial density profile, we can,
at present, not distinguish between a spatial model in which E+A galaxies
inhabit a shell around the cluster core, and one in which their distribution
increases toward the cluster core as is expected for the absorption-line
galaxies. Although a comparison of the respective projected radial distributions
of E+A and absorption-line galaxies suggests the former, a Kolmogorov-Smirnov
test finds them to be consistent with each other at the $1\sigma$ confidence
level. From our study of this cluster alone we are thus unable to address the
issue of whether E+A galaxies avoid the very cores of galaxy clusters as has
been suggested by other authors \citep{tran03,dres99}.

\subsection{Projected Density Distribution of Cluster Galaxies}

\begin{figure}[h]
\epsscale{1.0} \plotone{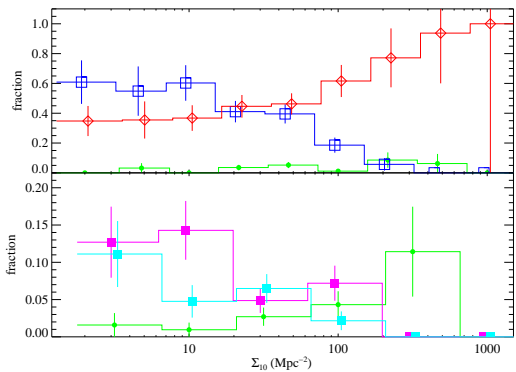}
\figcaption[f13.eps]{Fraction of galaxies of different spectral
  types as a function of projected local density. Symbols are as in
  Fig.~\ref{galxdistr}. Error bars assume Poisson statistics. The symbols for
  E+A galaxies and absorption-line galaxies are slightly shifted horizontally
  for clarity.\label{galxdensity}}
\end{figure}

One of the most important parameters for studies of galaxy evolution is the
density of the local environment. In Fig.~\ref{galxdensity} we plot the fraction
of galaxies of different types as a function of the projected local galaxy
density, $\Sigma_{10}$, as estimated from the sample of photometrically selected
cluster members (i.e., $z_{ph}=$0.507 to 0.580). The observed monotonic rise and
fall of the fractions of emission- and absorption-line galaxies (top panel) with
decreasing galaxy density, $\Sigma_{10}$, confirms our interpretation of the
trends already seen in Fig.~\ref{galxradial} and suggests strongly that the
projected density is a more fundamental physical parameter for the distribution
of the emission- and absorption-line galaxies than the distance to the cluster
center. This result is consistent with the relation between star-formation rate
and local density, as well as with the morphology-density relation, from studies
of other clusters at intermediate redshift \citep{dres80,pogg99,koda01}. Note,
however, how the fraction of either galaxy type is almost insensitive to the
density of the environment until the latter reaches a value typical of galaxy
groups ($\sim$ 20 Mpc$^{-2}$), at which point the divide between absorption- and
emission-line galaxies begins to open up dramatically \citep{koda01}.

The bottom panel of Fig.~\ref{galxdensity} emphasizes the stark contrast between
the environmental dependence of E+A galaxies and that of e(a) and e(b)
galaxies. The latter two galaxy types appear to, qualitatively, follow the trend
already seen for emission-line galaxies in general (which is not all that
surprising given that the e(a) and e(b) classification denotes subtypes of
emission-line galaxies), but with a possibly higher sensitivity to local density
already at values well below those characteristic of galaxy groups. E+A
galaxies, on the other hand, appear to follow the opposite trend of favoring
environments of intermediate to high density.

\begin{figure}[h]
\epsscale{1.0} \plotone{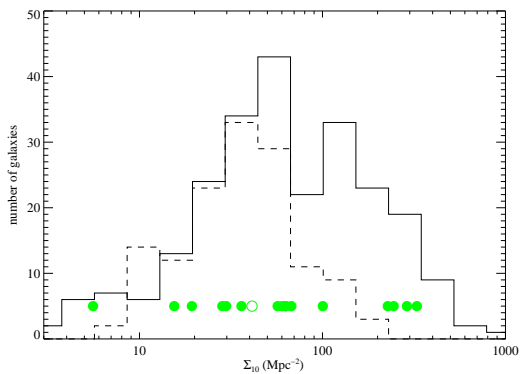}
\figcaption[paper_galdensity2pks.ps]{Histogram of projected local galaxy
  densities for Region A (solid line) and Region B (dashed line). See
  Fig.~\ref{plain} for a definition of these regions. The filled (open) green
  circles mark the densities at which we detect E+A galaxies within Region A
  (B).\label{galdens2pks}}
\end{figure}

The histogram of the $\Sigma_{10}$ values observed in Region A and Region B
shown in Fig.~\ref{galdens2pks} illustrates that the range of projected local
galaxy densities in Region B overlaps significantly with the one encountered in
the primary cluster region. Indeed, both emission- and absorption-line galaxies
show the same trends in either region (Figs.~\ref{galxradial},\ref{galxdensity}):
their fractions are mainly a function of $\Sigma_{10}$.  This is not true for
E+A galaxies: although the majority of E+A galaxies found in MACSJ0717.5+3745
are associated with $\Sigma_{10}$ values that are most commonly encountered in
Region B, only one out of our 17 E+A galaxies is in fact detected there. If
density were the determining factor in the creation of E+A galaxies, their
distribution with $\Sigma_{10}$ in Region A would lead to a prediction of 9
E+A galaxies in Region B -- a number that is inconsistent with the observed
count of one at the 3$\sigma$ significance level. We conclude that, at least
for the system under investigation here, the physical mechanism responsible for
the termination of star formation in this galaxy type must be a different one.

\section{Discussion}

The arguably most interesting result of our spectral analysis of the galaxy
population of the cluster/filament system MACSJ0717.5+3745 is the spatial
distribution of E+A galaxies, which may provide important clues about the
physical mechanism triggering the E+A transition phase between actively star
forming and passively evolving galaxies. Three aspects need to be considered in
this context: what is the origin of the galaxies that evolve into E+A galaxies,
what are the mechanisms that trigger the starburst, and what are the
mechanism(s) responsible for terminating it?

\subsection{E+A Progenitors}

A first clue about the origin of the E+A galaxies in MACSJ0717.5+3745 can be
obtained from their spatial distribution. If the galaxies that develop the E+A
spectral characteristics originated from the large-scale filament or the galaxy
groups embedded therein, one would expect E+A galaxies to be found
preferentially near the filament-cluster interface. According to
Fig.~\ref{galxdistr} this is not the case. Although the spatial distribution of
E+A galaxies exhibits a slight elongation in the direction of the (projected)
direction of infall along the filament, their location with respect to the
cluster core is consistent with a roughly spherical distribution which, in turn,
suggests an isotropic distribution of the progenitor population.

A different approach would be to attempt to directly identify the most likely
progenitors.  The strong H$_{\delta}$ absorption in the spectra of E+A galaxies
indicates that not only the starburst stopped both abruptly and recently (within
a few hundred millions years), but also that a very significant population of
young stars was thus generated (e.g.\ \citet{barger96}, and the review in
\citet{pogg04}). This is particularly true for this study since we have
intentionally selected the E+A galaxies with extreme Balmer absorption
features. Galaxies currently undergoing strong star formation i.e., the e(a) and
e(b) subtypes of emission-line galaxies, are thus natural candidates for the
role of progenitors of E+A galaxies, and should, to some extent, trace the
spatial distribution of the progenitor population.
Examining, again, Fig.~\ref{galxdistr} we see that both e(a) and e(b) galaxies
show a rather flat distribution which, however, prefers the general direction of
the filament and is, globally, distinctly un-isotropic with respect to the
cluster core. Although the distribution of e(a) and e(b) galaxies thus does not
rule out the hypothesis that the filament represents an important reservoir from
which E+A galaxies originate, the combined evidence appears to favor a picture
in which E+A galaxies evolve from a population of galaxies residing
predominantly in a halo surrounding the entire cluster+filament complex. The
characteristic density of this typical e(a)/e(b) environment is higher than that
of the infall region but lower than the one found in galaxy groups --- as a
result, these potential E+A progenitors are missing from the denser regions of
our study area as well as from the field.

Results from our complete survey of the galaxy population of the most distant
MACS clusters should allow us to address these issues with a much larger degree
of certainty.

\subsection{Source of Starburst Galaxies}

The next question is how the starbursts are triggered in cluster galaxies.  The
spatially fairly even distribution of e(a) and e(b) galaxies
Fig.~\ref{galxdistr}, which nonetheless shows a preference for the general area
of the filament, represents evidence that starbursts are triggered in
regions well outside the cluster, and in fact well outside the central regions
of the filament too. The densities typical of the environment in which e(a) and
e(b) galaxies predominantly reside can be estimated from both the spatial
distribution in the filament region (Fig.~\ref{galxdistr}) and our direct
measure of the projected galaxy density (Fig.~\ref{galxdensity}). Taking into
account the certainty of projection effects we arrive at \S5.5. This result that
star formation is triggered at radii approaching the virial radius and
environmental densities well below that of galaxy groups is qualitatively
consistent with the findings of \citet{koda01} and \citet{marc07}. 

\subsection{Star Formation Termination}

Although it is generally accepted that environmental effects other than galaxy
mergers are important to quench star formation in E+A galaxies in clusters
(\citet{tran03} and references therein), there has, so far, been little direct
observational evidence that would favor any particular physical mechanism.  From
our wide-field study of the spatial distribution of E+A and other types of
spectrally classified galaxies in MACSJ0717.5+3745, we believe to have found the
strongest evidence to date for ram-pressure stripping being the most effective,
and in fact possibly only, mechanism driving the rapid evolution of E+A galaxies
in clusters\footnote{Note that we refer here to the extreme E+A galaxies as
  defined by our strict criteria of Balmer line absorption. We do not rule out
  the possibility that other physical mechanisms play a significant role in the
  evolution of post-starburst galaxies with weaker Balmer absorption
  lines.}. This is because almost all of the E+A galaxies in the
MACSJ0717.5+3745 system are found to lie, in projection, near the cluster
core\footnote{Although spurious trends could conceivably be introduced into the
  observed spatial distribution if all or most of our E+A galaxies were in fact
  dust-enshrouded emission-line galaxies \citep{smail99,sato06}, it is difficult
  to imagine reasons for which any dust-extinction patterns should correlate
  this closely with cluster radius.} and within the ram-pressure stripping
radius $R_{rs}$. 
Simulations (e.g.\ \citep{tonn07}) show that ram-pressure stripping is the most
efficient mechanism to remove the gas in the galaxies, and, therefore, to
terminate star formation.  Furthermore, the complex substructure in both X-ray
emission (Fig.~\ref{xrayim}) and galaxy density (Fig.~\ref{galdenmap}) makes
MACSJ0717.5+3745 a perfect representative of a the kind of cluster merger
explored in the numerical simulation of \citet{fujita99}. The latter authors
find that the ram-pressure on galaxies near the cluster core increases
dramatically as clusters collide, and that cluster mergers also bring many blue
galaxies near the cluster center -- the combination of these two effects leads
to an increase in the fraction of post-starburst galaxies increases in the
cluster core.
The results of our study firmly rule out galaxy mergers as the primary driver
for the creation or evolution of E+A galaxies. If galaxy mergers were
responsible for the termination of star formation in E+A galaxies, we would
expect to see their distribution peak in regions where the galaxy densities and
relative velocities are most conducive to mergers, i.e., group-like environments
and the cluster infall region itself. This is clearly inconsistent with the
results presented in \S5.6 (Figs.~\ref{galxdistr} and \ref{galdens2pks}).  The
less violent galaxy-galaxy interaction mechanisms harassment and starvation are
ruled out too, not only because they cannot explain the observed concentration
of E+A galaxies near the regions of steeply increasing intra-cluster gas
density, but also because they act over effective time scales \citep{treu03} of
several billion years and are thus too slow to sharply terminate the star
formation in E+A galaxies. By contrast, the time \citep{treu03} needed for a
galaxy to fall from the virial radius, where star formation would be triggered,
to the ram-pressure stripping radius, where the star formation would be
terminated again essentially instantaneously, and then on to the cluster core
($\sim$ 1Gyr) is well matched to the lifetime of the E+A phase ($\sim$ 1Gyr). We
note though that galaxy mergers are not ruled out in terms of effective time
scale (a few hundreds million years) as a mechanism for the creation of E+A
galaxies in general. In the field mergers appear indeed to be the dominant
mechanism \citep{zabl96,yang04,goto03, tran04}.  This dichotomy between the
origin of E+A galaxies in low- and high-density environments has previously been
suggested by \citet{tran03} and \citet{tran04}.

\subsection{Comparison to Other Clusters} 

\begin{figure}[t]
\epsscale{1.0} 
\plotone{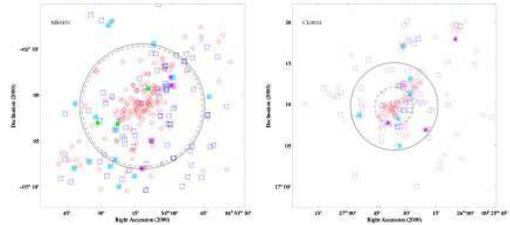}
\figcaption[paper_MS0451.eps]{Spatial distribution of different galaxy types in
  MS0451.6--0305 and Cl0024+16 based on data from \citet{mora07}. The symbols
  are the same as in Fig.~\ref{galxdistr}. The solid circle has radius
  $R_{vir}$, whereas the dashed circle marks $R_{rs}$ as shown in Figure 1 of
  \citet{mora07}. [The full resolution figure will be published in ApJ]
\label{Moran}}
\end{figure}

To test whether our conclusions extend beyond MACSJ0717.5+3745, we investigate
the distribution of E+A galaxies in the only two other clusters at intermediate
redshift for which spectroscopic data of comparable quality are available over a
similarly large field of view, namely MS0451.6--0305 and Cl0024+16\footnote{EW
  measurements for these systems were obtained from
  http://www.astro.caltech.edu/$\sim$smm/clusters/.}\citep{treu03,mora07}. Although
the details of their survey in \citet{mora07}, such as the selection criteria,
the magnitude limit, the instrumental setup, and the technical details of the EW
measurements\footnote{\citet{mora07} adopt the Lick system\citep{worthey94} for
  their EW measurements; we perform our spectral classification using their
  $H{\delta_A}$ and $H{\gamma_A}$ values. Although use of the Lick-system
  results will affect our classification to some extent, we expect the
  differences to be insignificant comparing to other systematic uncertainties.},
are different from those of our work, a qualitative comparison should still
be permitted. In addition, we attempt a comparison between MACSJ0717.5+3745 and
a similarly massive cluster in the local universe, to test whether the spatial
distribution of E+A galaxies is consistent. This would imply that, regardless of
any redshift evolution in the relative prevalence of galaxies of a given
spectroscopic type, ram-pressure stripping is the sole physical mechanism
driving the evolution of E+A galaxies in clusters at all epochs. Since E+A
galaxies are rare in local clusters we are left with a single system featuring
enough E+A galaxies for a statistically meaningful comparison, the Coma
cluster\citep{pogg04}.
  
MS0451.6--0305 ($z=0.54$, L$_{x} = 17~10^{44}$ergs~s$^{-1}$, T$_x = 7$~keV,
Gioia \& Luppino 1994, Ebeling et al.\ 2007) is comparable to MACSJ0717.5+3745
in that it is one of the most X-ray luminous and best-studied clusters at $z\sim
0.5$. Unlike MS0451.6--0305 and MACSJ0717.5+3745, Cl0024+16 ($z=0.39$)
is an optically selected cluster and much less X-ray luminous (L$_{x} =
2.9~10^{44}$ergs~s$^{-1}$, T$_{x}=3.5$~keV, Zhang et al.\ 2005). We nonetheless
include here for comparison purposes to investigate the relevance of cluster
properties such as total mass, dynamical state, and core density. The Coma
cluster, our local reference, is the most massive cluster at $z<0.05$ (T$_{x}=9.0$~keV (Donnelly et al. 1999).

As shown in Fig.~\ref{Moran}, E+A galaxies are rare in both MS0451.6--0305
($1^{+1.1}_{-0.6}$\%) and Cl0024+16 ($0^{+0.7}_{-0}$\%) compared to
MACSJ0717.5+3745 ($3.2^{+1.0}_{-0.8}$\%), although the uncertainties are
large. However, their spatial distribution (all within $R_{rs}$) is consistent
with what we observe in MACSJ0717.5+3742. In addition, the fact that the E+A
galaxies in MS0451.6--0305 avoid the dense cluster core lends support to the
hypothesis (see \S5.5) that the true three-dimensional spatial distribution of
this type of galaxies resembles a shell inside the ram-pressure stripping
radius. No E+A galaxy at all is found in Cl0024+16. However, as reflected in its
much smaller ram-pressure stripping radius, this cluster is much less massive
than either of MS0451.6--0305 and MACSJ0717.5+3745. In fact, the density of this
cluster's core region is still likely to be overestimated because of the unusual
geometric and dynamical properties of Cl0024+16, which is known to be a
high-velocity merger along our line of sight \citep{czoske02}.

Our results for MACSJ0717.5+3742 are 
consistent with the distribution of e(a)/e(b) galaxies in MS0451.6--0305 and
Cl0024+16, too. In either cluster, these starburst candidates are distributed much
more widely than the E+A or absorption-line population.

The spatial distribution of E+A galaxies in Coma \citep{pogg04} is also
consistent with our result\footnote{The study of \citet{pogg04} addresses
  properties of k+a galaxies selected based on $H_{\delta}$ and $[OII]$;
  although this definition is different from ours, the selected galaxy type is
  similar to our E+A galaxies.}. \citet{pogg04} separate E+A galaxies into two
categories according to their color. Based on the relative position of the red
sequence and the E+A galaxies in Fig.~\ref{cc}, the E+A galaxies in
MACSJ0717.7+3745 mostly belong to the class of "blue E+A" galaxies, which
\citet{pogg04} find to be concentrated near the core of Coma. A similar
sub-population is selected by focusing on the "extreme" E+A galaxies with
$H_{\delta} > 6$\AA (see also \S4.3). Applying this criterion to the k+a sample
of \citet{pogg04} we again find most of them (7 out of 9) to be located near the
cluster center but avoiding the very core. Finally, \citet{pogg04} speculate
that the distribution of these youngest E+A galaxies in Coma traces the X-ray
substructure, suggesting that the origin of these extreme E+A galaxies may be
related to merger-induced shocks in the cluster gas. While, qualitatively, this
picture is not in conflict with our findings for MACSJ0717.7+3745, a known major
merger (Fig.~\ref{xrayim} shows the complex X-ray morphology of the cluster
core), this hypothesis would make the absence of E+A galaxies in Cl0024+16 even
harder to explain.

\section{Conclusion}

MACSJ0717.5+3745 is a very X-ray luminous cluster at $z=0.55$ which shows
complex sub-cluster and filamentary structure in both X-ray surface brightness
and galaxy density. Along the filament, we detect a significant offset in the
average redshift of galaxies, corresponding to $\sim 630$km~s$^{-1}$ in
velocity, as well as a decrease in velocity dispersion from the core of the
primary cluster to the end of filament. These variations are likely a
combination of spatial (Hubble flow) and kinematic effects (peculiar velocities)
in this highly disturbed, merging system; a detailed analysis and discussion of
the three-dimensional geometry and dynamics of MACSJ0717.5+3745 will be
presented in a forthcoming paper.

Using the spectroscopic data obtained by us, primarily with the DEIMOS
spectrograph on the KECK-II telescope, we perform an extensive analysis of the
spatial distribution of galaxies of different spectroscopic types, from the
cluster core to the cluster outskirts and the connected linear filament. The
spectroscopic sample is selected using blue and red color cuts defined relative
to the empirical cluster red sequence in the V and R$_{c}$ bands. The final
spectroscopic catalog is more than $70\%$ complete at R$_{c} < 21.75$. The applied
blue color cut causes the number of emission-line galaxies in this structure, in
particular the starburst galaxies, to be underestimated -- our analysis of the
spatial distribution of the cluster members of various spectroscopic types should,
however, not be affected.

We classify cluster members into emission-line, absorption-line, and E+A
galaxies using the strength of the [OII] and H$_{\beta}$ emission lines, as well
as the absorption features H$_{\delta}$ and H$_{\gamma}$. The emission-line
galaxies are further categorized into the subtypes e(a), e(b), and e(c),
depending on the strengths of the [OII] emission- and H$_{\delta}$
absorption-line features. We emphasize that in order to eliminate contamination
from absorption- or emission-line galaxies, we adopt a very strict criterion for
our definition of E+A galaxies, namely $\frac{(H_{\delta}+H_{\gamma})}{2} >
6$\AA, and also require the absence of [OII] and H$_{\beta}$ emission.  The
location of these E+A galaxies in color-magnitude and color-color diagrams
confirms that they represent the transition phase between emission-line
galaxies, in particular the e(a) and e(b) sub-types, and absorption-line
galaxies.

The main result of this study is that E+A galaxies are found to reside almost
exclusively within the ram-pressure radius of the cluster. Since E+A galaxies
are not observed in significant numbers in other regions of this system where
the local galaxy density is similarly high, and since the relatively high galaxy
velocities near the ram-pressure radius are unfavourable for galaxy mergers, we
conclude that galaxy-gas interactions, rather than galaxy mergers, are likely
the dominant mechanism for the evolution of the E+A galaxies, at least in
clusters at intermediate redshift. Our interpretation that the star formation in
these galaxies is terminated because the gas reservoir is being stripped away by
the intra-cluster medium as galaxies fall into the cluster core is consistent
with the results obtained for other two clusters, MS0451.6--0305 and
Cl0024+16. E+A galaxies are only found within the ram-pressure radius of
MS0451.6--0305, and no E+A galaxies at all are found in Cl0024+16, a low-mass
system featuring a very small ram-pressure stripping radius.  The large number
of E+A galaxies around the core of MACSJ0717.7+3745, compared to the similarly
X-ray luminous cluster MS0451.6--0305, can be naturally explained by the
significant merger activity in the latter system, an effect that is consistent
with the results of numerical simulations.

Although the spatial distribution of E+A galaxies in MACSJ0717.7+3745 shows
clear evidence of environmental effects on galaxy evolution in clusters, many
issues still need further investigation. Specifically, we need to better
understand the role of cluster mergers, and what exactly triggers starbursts as
galaxies are falling into the cluster. Both of these questions can be addressed
by extending this work to the full sample of the 12 most distant MACS clusters,
thereby covering systems of very different dynamical states and a wide range of
large-scale environments. For MACSJ0717.7+3745, accurate star-formation rates
measured from infrared data obtained with Spitzer will help greatly in this
regard. In addition we aim to use galaxy morphologies and deep X-ray
observations to study in detail the gas-galaxy interactions which we suspect to
be responsible for the termination of star formation activity and thus the
creation of E+A galaxies in clusters.\\ \\

We thank Leif Wilden and Glenn Morris of Stanford University for assistance with
the photometric calibration of our SuprimeCam data for MACSJ0717.7+3745 and the
anonymous referee for advice and suggestions that helped to improve this
paper. Financial support for this work was provided by the National Aeronautics
and Space Administration through Chandra Award Number GO3-4168X issued by the
Chandra X-ray Observatory Center, which is operated by the Smithsonian
Astrophysical Observatory for and on behalf of the National Aeronautics Space
Administration under contract NAS8-03060. This research has made use of software
provided by the Chandra X-ray Center (CXC) in the application packages CIAO,
ChIPS, and Sherpa. The analysis pipeline used to reduce the DEIMOS data was
developed at UC Berkeley with support from NSF grant AST-0071048.

\end{document}